\documentclass[review]{elsarticle}
\biboptions{sort&compress} 

\usepackage{epsfig,amsmath,amssymb}
\usepackage{enumerate}
\usepackage{subfigure}
\usepackage{graphicx} 
\usepackage{epstopdf}
\usepackage{booktabs}
\usepackage{multirow}
\usepackage{xcolor}
\usepackage{appendix}
\usepackage{threeparttable}
\journal{arXiv}

\usepackage{amsmath,amssymb}
\usepackage{subfigure,algorithm,algorithmic}

\makeatletter
\newenvironment{breakablealgorithm}
{
	\begin{center}
		\refstepcounter{algorithm}
		\hrule height.8pt depth0pt \kern2pt
		\renewcommand{\caption}[2][\relax]{
			{\raggedright\textbf{\ALG@name~\thealgorithm} ##2\par}%
			\ifx\relax##1\relax 
			\addcontentsline{loa}{algorithm}{\protect\numberline{\thealgorithm}##2}%
			\else 
			\addcontentsline{loa}{algorithm}{\protect\numberline{\thealgorithm}##1}%
			\fi
			\kern2pt\hrule\kern2pt
		}
	}{
		\kern2pt\hrule\relax
	\end{center}
}
\makeatother

\usepackage{amsmath}

        \makeatletter
        \def\fps@eqnfloat{!t}
        \def\ftype@eqnfloat{4}
        
        \newenvironment{eqnfloat*}
               {\@dblfloat{eqnfloat}}
               {\end@dblfloat}
        \makeatother
\makeatletter
\let\oldabs\abs
\def\abs{\@ifstar{\oldabs}{\oldabs*}}
\let\oldnorm\norm
\def\norm{\@ifstar{\oldnorm}{\oldnorm*}}
\makeatother

\renewenvironment{thebibliography}[1]{%
	\begin{oldthebibliography}{#1}%
		\setlength{\parskip}{0ex}%
		\setlength{\itemsep}{0ex}%
	}%
	{%
	\end{oldthebibliography}%
}%

\begin{document}
\begin{frontmatter}
\title{Waveform Optimization with SINR Criteria for FDA Radar in the Presence of Signal-Dependent Mainlobe Interference \tnoteref{t1}}
\tnotetext[t1]{The work has been supported in part by the National Natural Science Foundation of China under grant 62171092 and in part by the Swedish SRA ESSENCE (grant no. 2020 6:2).}

%
%
%
%
%
%

\author[addr]{Wenkai Jia\corref{cor}}
\address[addr]{School of Information and Communication Engineering, University of Electronic Science and Technology of China, 611731, Chengdu, P. R. China}
\ead{wenkai.jia@matstat.lu.se}

\author[addrr]{Andreas Jakobsson}
\address[addrr]{Department of Mathematical Statistics, Center for Mathematical Sciences, Lund University, 22100, Lund, Sweden}
\cortext[cor]{Corresponding author.}
\ead{andreas.jakobsson@matstat.lu.se}

\author[addr]{Wen-Qin Wang}
\ead{wqwang@uestc.edu.cn}

%
%

\begin{abstract}
In this paper, we focus on the design of the transmit waveforms of a frequency diverse array (FDA) in order to improve the output signal-to-interference-plus-noise ratio (SINR) in the presence of signal-dependent mainlobe interference. Since the classical multi-carrier matched filtering-based FDA receiver cannot effectively utilize the waveform diversity of FDA, a novel FDA receiver framework based on multi-channel mixing and low pass filtering is developed to keep the separation of the transmit waveform at the receiver side, while preserving the FDA range-controllable degrees of freedom. Furthermore, a range-angle minimum variance distortionless response beamforming technique is introduced to synthesize receiver filter weights with the ability to suppress a possible signal-dependent mainlobe interference. The resulting FDA transmit waveform design problem is initially formulated as an optimization problem consisting of a non-convex objective function and multiple non-convex constraints. To efficiently solve this, we introduce two algorithms, one based on a signal relaxation technique, and the other based on the majorization minimization technique. The preferable performance of the proposed multi-channel low pass filtering receiver and the optimized transmit waveforms is illustrated using numerical simulations, indicating that the resulting FDA system is not only able to effectively suppress mainlobe interference, but also to yield estimates with a higher SINR than the FDA system without waveform optimization.
\end{abstract}
\end{frontmatter}

\section{Introduction}
\noindent
Increasing the controllable degrees of freedom (DoF) of a radar system
enables an improved ability to collect information about possible targets and their environment \cite{1597550,1589439,7771665}. 
An example of a way to create such an improvement is the uniformly phased array (UPA) radar system, which realizes a flexible beam scanning by the means of electronically controlling the phase and amplitude of the transmitted waveform. 
The introduction of multiple transmitters and receivers enables the use of multiple-input multiple-output (MIMO) radar systems \cite{4350230}, which may transmit multiple probing signals, each of which may be optimized to obtain a desired beampattern \cite{4276989,6649991,8706630}.
Such a system may be further refined by the use of a frequency diverse array (FDA) \cite{1631800,7740083}, being capable of generating a range-angle-time-dependent transmit beampattern by the use of a small frequency increment across the array elements \cite{1631858}.

Theoretically, the frequency difference between FDA transmitting elements makes the range information of different targets manifest in the form of phase differences of FDA echoes at the receiving elements. 
Most importantly, this enables a range information that is different from that determined by the time delay used in the traditional radar system. Indeed, similar to the spatial beamforming \cite{1165054}, the FDA supports target discrimination by the use of range beamforming \cite{8321488,6376087,7088582}. Therefore, frequency diversity introduces an additional range-controllable DoF to the FDA system.
It has been shown that FDA offers preferable performance in joint angle and range localization \cite{6630081,6737322,7084678}, radio frequency (RF) stealth \cite{7422108,7362565}, and in low probability of intercept (LPI) \cite{9440812} as compared to its UPA or MIMO counterparts.

However, complex electromagnetic environment creates challenges for the target detection and tracking of radar systems. For example, a jamming system based on digital radio frequency memory (DRFM) technology \cite{neng1995survey,olivier2011design} can replicate the waveform characteristics and intra-pulse information of the radar transmit signal, and thereby deceive the radar system in the joint time-frequency domain. Hence, the capability of the system's jamming suppression becomes a key indicator for evaluating the radar's performance \cite{richards2010principles}. 
Owing to its importance for the system's efficiency,
various methods to mitigate deceptive jammers have been studied in the literature (see, e.g., \cite{skaria2019interference,haykin2006cognitive}).
Even though such schemes are efficient, signal-dependent mainlobe interference is still highly problematic and will significantly degrade the performance.
In MIMO radar, if the range of the interference is known, one possible solution is to use a weighted integrated sidelobe level (ISL) minimization approach \cite{he2009designing,song2016sequence,li2017fast}. In such an approach, just some of the lags of the sidelobes are minimized, not the full range. However, this method is not suitable for the situation where the interference and the target are located in the same range bin.
In contrast, by exploring the DoF in the range domain, several mitigation methods for these mainlobe interferences have been proposed for FDA systems, including the projection matrix method \cite{8902536}, as well as data-dependent \cite{8448961} and data-independent beamforming \cite{9161264} methods.

It is worth noting that these methods were all developed under the assumption of the multi-carrier matched filtering-based receiver proposed in \cite{8074796}.
It is well known that the matched filter maximizes the signal-to-noise ratio (SNR) in the presence of additive white noise, but notably not the signal-to-clutter ratio (SCR) \cite{6104176,4567663}.
As the overall performance of the system will depend on the output signal-to-interference-plus-noise ratio (SINR), the performance loss of the matched filter-based FDA receiving systems is rarely highlighted.
This loss may be alleviated by exploring the additional DoF resulting from an efficient transmit waveform design.
The problem of transmit waveform design in MIMO radar systems has been extensively studied \cite{6649991,7470499,8141978}, whereas the extension of such systems to also allow for FDA transmit waveform design, does not appear available in the open literature.

Different from current FDA studies, which mainly focuses on designing receiving weights to effectively suppress mainlobe interference, in this work, we aim to improve the output SINR by instead designing the FDA transmit waveform.
First, by exploring the spectral characteristics of the FDA transmit waveform, we design an FDA receiver that can effectively receive FDA echoes and maximally separate the transmitted waveforms.
Next, the corresponding receiver filter weights are devised to allow for the presence of interference by utilizing the minimum variance distortionless response beamforming (MVDR) techniques.
Finally, the problem of FDA transmit waveform design is formulated as a non-convex multi-constrained optimization problem.
To facilitate the waveform design, we devise two iterative algorithms. Specifically, at each iteration of the first algorithm, with the help of the previous iteration results, the non-convex objective function consisting of signal-dependent interferences is first approximated to a quadratic convex objective function with respect to the transmit waveform and the signal-independent interferences. Next, an auxiliary variable is introduced so that the obtained approximative problem can be well incorporated into the alternating direction method of multipliers (ADMM) framework \cite{boyd2011distributed}.
Furthermore, we introduce a second algorithm inspired by the majorization-minimization (MM) method that does not directly optimize the objective function but instead optimizes a simple surrogate function that locally approximates the objective function with their difference being minimized at the current point \cite{7547360}. The resulting MM-ADMM method optimizes this surrogate function and solves the resulting problem by employing a set of auxiliary variables.
Computer simulations are provided to demonstrate the effectiveness of the proposed techniques with respect to the range-angle beampattern, the SINR performance, as well as the pulse compression property.

The remaining sections are organized as follows.
In Section \ref{sec:a}, 
we introduce the designed FDA receiver and construct the novel FDA receiver signal model considering the radar range gate, as well as derive the optimal receive filter weights based on the MVDR principle.
Taking into account various practical waveform constraints, maximizing the waveform-dependent SINR is discussed in Section \ref{sec:b}.
Section \ref{sec:c} presents two optimization algorithms solving the resulting optimization problem. 
Section \ref{sec:d} provides numerical results illustrating the achievable 
performance gain. Finally, conclusions are drawn in Section \ref{sec:e}.

\emph{Notations:} Bold lowercase letter ${\mathbf{a}}$ and uppercase letter ${\mathbf{A}}$ represent vectors and matrices, respectively.
The transpose, conjugate, and conjugate transpose operators are denoted by ${\left(  \cdot  \right)^T}$, ${\left(  \cdot  \right)^c}$, and ${\left(  \cdot  \right)^H}$, respectively.
For a matrix ${\mathbf{A}}$, we use ${\mathbf{A}}\left( {i,j} \right)$ to indicate its $i$th row and $j$th column element, whereas $\operatorname{Tr} \left\{ {\mathbf{A}} \right\}$ denotes the trace of ${\mathbf{A}}$.
The symbol $\operatorname{vec} \left\{ {\mathbf{A}} \right\}$ is the vectorization operator, $\operatorname{diag} \left\{ {\mathbf{a}} \right\}$ denotes the diagonal matrix with the diagonal entries formed by ${\mathbf{a}}$, and ${{\left\| \mathbf{a} \right\|}_{\infty }}$ and ${{\left\| \mathbf{a} \right\|}_{2}}$ represent the infinity norm and Euclidean norm of ${\mathbf{a}}$, respectively.
Furthermore, ${{\mathbf{I}}_M}$ and ${{\mathbf{1}}_M}$ denote the $M$-dimensional identity and all-ones matrix, respectively, whereas
${{\mathbf{e}}_{j}}$ indicates a vector with the $j$th element being $1$ and the others $0$.
The sets of $M \times M$ real-valued and complex-valued matrices are denoted ${\mathbb{R}^{M \times M}}$ and ${\mathbb{C}^{M \times M}}$, respectively.
Finally, the Hadamard and Kronecker matrix products are written as $ \odot $ and $ \otimes$, respectively.

\section{Signal model}\label{sec:a}
\subsection{Transmit and echo signal model}
\noindent
Consider a uniform linear FDA with ${N_T}$ equispaced transmit antennas with uniform frequency increment $\Delta f$.
Then, the transmit signal $x\left( t \right)$ may be expressed as \cite{8074796} 
\begin{equation}
x\left( t \right)=\sum\limits_{m=1}^{{{N}_{T}}}{\left\{ {{s}_{m}}\left( t \right){{e}^{-j2\pi \left( {{f}_{c}}+\left( m-1 \right)\Delta f \right)t}} \right\}}
\end{equation}
where ${s_m}\left( t \right)$ denotes the transmit waveform that excites the $m$th array element and $f_c$ represents the reference carrier frequency.
Under the narrow-band assumption, the return signal reflected by a far-field target with range-angle pair $\left( {{r}_{t}},{{\theta }_{t}} \right)$ can be expressed as \cite{8074796}
\begin{equation}\label{eq:ynt}
\begin{aligned}
{{y}_{n}}\left( t \right)&=\xi \left( {{r}_{t}},{{\theta }_{t}} \right)\sum\limits_{m=1}^{{{N}_{T}}}{\left\{ {{s}_{m}}\left( t-{{\tau }_{m,n}} \right)\varphi \left( t-{{\tau }_{m,n}} \right) \right\}} \\ 
& \approx \varsigma \left( {{r}_{t}},{{\theta }_{t}} \right)\sum\limits_{m=1}^{{{N}_{T}}}{\left\{ \begin{aligned}
	& {{s}_{m}}\left( t-{2{{r}_{t}}}/{c}\; \right){{e}^{-j2\pi \left[ {{f}_{c}}+\left( m-1 \right)\Delta f \right]t}} \\ 
	& {{e}^{{j2\pi \left( m-1 \right)\Delta f2{{r}_{t}}}/{c}\;}}\cdot {{{\bar{\varphi }}}_{{{d}_{t}}}}\cdot {{{\bar{\varphi }}}_{{{d}_{r}}}} \\ 
	\end{aligned} \right\}} \\ 
\end{aligned}
\end{equation}
for $n=1,2,...,{{N}_{R}}$ with ${N_R}$ representing the number of receive antennas, where $\varphi \left( t-{{\tau }_{m,n}} \right)={{e}^{-j2\pi \left[ {{f}_{c}}+\left( m-1 \right)\Delta f \right]\left( t-{{\tau }_{m,n}} \right)}}$, ${{{\bar{\varphi }}}_{d}}={{e}^{{-j2\pi \left[ {{f}_{c}}+\left( m-1 \right)\Delta f \right]\left( m-1 \right)d\sin {{\theta }_{t}}}/{c}\;}}$, and
$\varsigma \left( {{r_t},{\theta _t}} \right) = \xi \left( {{r_t},{\theta _t}} \right){e^{j2\pi {f_c}\frac{{2r_t}}{c}}}$, with $\xi \left( {{r_t},{\theta _t}} \right)$ being the complex amplitudes of the target. ${\tau _{m,n}} = \frac{{2{r_t} - \left( {m - 1} \right)d_t\sin {\theta _t} - \left( {n - 1} \right)d_r\sin {\theta _t}}}{c}$ represents the round-trip delay from the $m$th transmit and the $n$th receive antennas, where $d_t$ and $d_r$ denote the transmit and receive inter-element spacing, respectively, with $c$ being the speed of light.
Under the assumption that $\Delta f \ll {f_c}$, \eqref{eq:ynt} can be concisely written as \cite{6737322}
\begin{equation}\label{eq:sbt}
\begin{aligned}
& {{y}_{n}}\left( t \right)\approx \varsigma \left( {{r}_{t}},{{\theta }_{t}} \right){{e}^{-j2\pi \frac{\left( n-1 \right){{d}_{r}}\sin {{\theta }_{t}}}{\lambda }}}\mathbf{a}_{\operatorname{T}}^{T}\left( {{r}_{t}},{{\theta }_{t}} \right)\mathbf{\tilde{s}}\left( t \right) \\ 
\end{aligned}
\end{equation}
where 
\begin{subequations}
	\begin{equation}
	\kern -29pt \mathbf{\tilde{s}}\left( t \right)=\mathbf{s}\left( t-\frac{2{{r}_{t}}}{c} \right)\odot \mathbf{e}\left( t \right)
	\end{equation}
	\begin{equation}
	\kern 2pt \mathbf{s}\left( t \right)={{\left[ {{s}_{1}}\left( t \right),{{s}_{2}}\left( t \right),...,{{s}_{{{N}_{T}}}}\left( t \right) \right]}^{T}}
	\end{equation}
	\begin{equation}
	\kern 14pt \mathbf{e}\left( t \right)={{\left[ \begin{aligned}
			& {{e}^{-j2\pi {{f}_{c}}t}},{{e}^{-j2\pi \left( {{f}_{c}}+\Delta f \right)t}} \\ 
			& ,...,{{e}^{-j2\pi \left( {{f}_{c}}+\left( {{N}_{T}}-1 \right)\Delta f \right)t}} \\ 
			\end{aligned} \right]}^{T}}
	\end{equation}
	\begin{equation}
	\kern 12pt {{\mathbf{a}}_{\operatorname{T}}}\left( {{r}_{t}},{{\theta }_{t}} \right)={{\left[ \begin{aligned}
			& 1,{{e}^{j2\pi \left( \frac{2\Delta f{{r}_{t}}}{c}-\frac{{{d}_{t}}\sin {{\theta }_{t}}}{\lambda } \right)}} \\ 
			& ,...,{{e}^{j2\pi \left( {{N}_{T}}-1 \right)\left( \frac{2\Delta f{{r}_{t}}}{c}-\frac{{{d}_{t}}\sin {{\theta }_{t}}}{\lambda } \right)}} \\ 
			\end{aligned} \right]}^{T}}
	\end{equation}
\end{subequations}
where $\mathbf{s}\left( t \right)$ and $\mathbf{e}\left( t \right)$ denote the $N_T$-dimensional transmit waveform and carrier vector, respectively, whereas ${{\mathbf{a}}_{\operatorname{T}}}\left( r_t,\theta_t  \right)$ denotes the FDA transmit steering vector and $\lambda =\frac{c}{{{f}_{c}}}$ is the reference wavelength.

\subsection{Receiver design}
\begin{figure*}[t]
	\centering
	\includegraphics[width=0.9\textwidth]{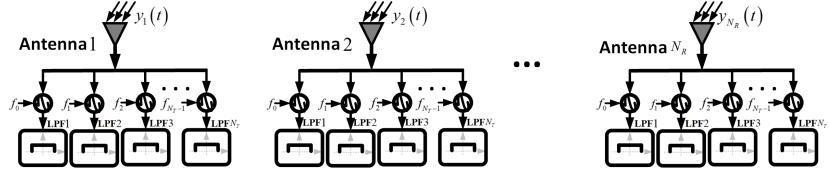}
	\caption{The designed FDA receiver.}\label{fig:111}
\end{figure*}

\noindent
Note that in \eqref{eq:sbt}, the transmit waveform vector and the carrier vector are coupled, which is different from UPA and MIMO systems.
Combined with the requirement of delay orthogonality \cite{9266663}, the multi-carrier matched filtering-based receiver can effectively detect the FDA echo signal \cite{8074796}, but is unable to exploit the full dynamics of the transmit waveform.
To allow for this flexibility, we here introduce a novel
FDA receiver framework based on multi-channel mixing and low pass filtering, which demodulates the returned FDA signals and then decouples the transmit waveform, as illustrated in Figure \ref{fig:111}.
The echo signal ${{y}_{n}}\left( t \right)$ is then down-converted by a multi-channel mixer with local carrier frequencies $\left\{ {{f}_{m}}={{f}_{c}}+m\Delta f \right\}_{m=0}^{{{N}_{T}}-1}$ followed by a low pass filter (LPF) to process the returned FDA signal sequentially.

Thus, the proposed receivers perform frequency band separation of multi-carrier FDA echo signals (consisting of multiple transmit waveforms located in different frequency bands) by low pass filtering.
As a result, these structures are only suitable for cases when the waveform bandwidth of each transmit antenna is smaller than the frequency increment.
We proceed to examine the advantages of the designed receiver in utilizing waveform diversity.

\subsection{Receive signal model}
\begin{figure}[t]
	\centering
	\includegraphics[width=0.47\textwidth]{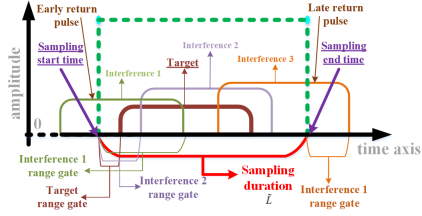}
	\caption{The diagram of the received signals.}\label{fig:222}
\end{figure}
\noindent The measured signal will clearly depend on the used receiver structure; for the receiver in Figure \ref{fig:111}, the filtered output in the $n$th receive antenna can be expressed as \cite{8074796}
\begin{equation}
\begin{array}{l}
{{\bf{r}}_n}\left( t \right) \\ = \left[ {{\mathbf{e}}^{c}}\left( t \right)\cdot {{y}_{n}}\left( t \right) \right]*\mathbf{h}\left( t \right)\\
\kern 0pt = \varsigma \left( {{r_t},{\theta _t}} \right){e^{ - j2\pi \frac{{\left( {n - 1} \right){d_r}\sin \theta }}{\lambda }}}{\mathop{\rm diag}\nolimits} \left\{ {{{\bf{a}}_{\mathop{\rm T}\nolimits} }\left( {{r_t},{\theta _t}} \right)} \right\}
\cdot {\bf{s}}\left( {t - \frac{{2{r_t}}}{c}} \right)\\
\end{array}
\end{equation}
for $n = 1,2,...,{N_R}$, where ${\bf{h}}\left( t \right) = {\left[ {{h_1}\left( t \right),{h_2}\left( t \right),...,{h_{{N_T}}}\left( t \right)} \right]^T}$ represents the LPF transfer function vector, where the cutoff frequency ${f_{m,lp}}$ of the $m$th LPF transfer function ${h_m}\left( t \right)$ satisfies
\begin{equation}
{B_{{s_m}\left( t \right)}} <  = {f_{m,lp}} <  = \Delta f, \kern 16pt m = 1,2,...,{N_T}
\end{equation}
where ${{B}_{{{s}_{m}}\left( t \right)}}$ denotes the bandwidth of the $m$th transmit waveform and the subscript $_{lp}$ denotes the LPF.
It should be noted that the constraint that the frequency increment is larger than the bandwidth of the baseband signal produces an FDA signal with non-overlapping spectra. This implies the function $\mathbf{h}\left( t \right)$ is able to suppress the signal outside the relevant band.
The received ${{N}_{T}}{{N}_{R}}\tilde{L}$-dimensional complex-valued FDA vector can thus be written as
\begin{equation}
\kern -1pt \begin{aligned}
{{{\mathbf{\hat{r}}}}_{\text{FDA}}}&=\operatorname{vec}\left\{ {{{\mathbf{\hat{R}}}}_{\text{FDA}}} \right\} \\ 
& =\varsigma \left( {{r}_{t}},{{\theta }_{t}} \right)\mathbf{A}\left( {{r}_{t}},{{\theta }_{t}} \right)\mathbf{s}+\sum\limits_{i=1}^{\Upsilon }{\left\{ \varsigma \left( {{r}_{i}},{{\theta }_{i}} \right)\mathbf{A}\left( {{r}_{i}},{{\theta }_{i}} \right)\mathbf{s} \right\}}+\mathbf{n} \\ 
\end{aligned}
\end{equation}
with $\mathbf{s}=\text{vec}\left\{ \mathbf{S} \right\}$, $\mathbf{n}=\text{vec}\left\{ {\mathbf{\bar{N}}} \right\}$, and 
$\mathbf{A}\left( r,\theta  \right)=\mathbf{K}{{\left( {{l}^{r}} \right)}^{T}}\otimes \left[ {{\mathbf{a}}_{\text{R}}}\left( \theta  \right)\otimes \text{diag}\left\{ {{\mathbf{a}}_{\text{T}}}\left( r,\theta  \right) \right\} \right]$,
where
$\mathbf{s}\in {{\mathbb{C}}^{{{N}_{T}}L\times 1}}$,
$\mathbf{n}\in {{\mathbb{C}}^{{{N}_{T}}{{N}_{R}}\tilde{L}\times 1}}$,
and
$\mathbf{A}\left( r,\theta  \right)\in {{\mathbb{C}}^{{{N}_{T}}{{N}_{R}}\tilde{L}\times {{N}_{T}}L}}$.
$ {{{\mathbf{\hat{R}}}}_{\text{FDA}}}$, $\mathbf{S}$, ${\mathbf{\bar{N}}}$, and $\mathbf{K}{{\left( {{l}^{r}} \right)}^{T}}$ are defined in \ref{sec:appenRx}.

\subsection{Receive filter weights}
\noindent
In order to facilitate the subsequent processing, we rewrite the two steering vectors ${{\bf{a}}_{\mathop{\rm T}\nolimits} }\left( {{r},{\theta}} \right)$ and ${{\mathbf{a}}_{\text{R}}}\left( {{\theta }_{t}} \right)$ as
\begin{subequations}
	\begin{equation}
	\kern 0pt {{\bf{u}}_{\mathop{\rm T}\nolimits} }\left( {{f_{t,t}} } \right) =  {\left[ {1,{e^{j2\pi {f_{t,t}}}},...,{e^{j2\pi \left( {{N_T} - 1} \right){f_{t,t}}}}} \right]^T}
	\end{equation}
	\begin{equation}
	{{\bf{u}}_R}\left( {{f_{r,t}} } \right) = {\left[ {1,{e^{j2\pi {f_{r,t}}}},...,{e^{j2\pi \left( {{N_R} - 1} \right){f_{r,t}}}}} \right]^T}
	\end{equation}
\end{subequations}
where ${f_{t,t}} = \frac{{2\Delta f{r_t}}}{c} - \frac{{{d_t}\sin \theta_t }}{\lambda },{f_{r,t}} =  - \frac{{{d_r}\sin \theta_t }}{\lambda }$
represent the transmit and receive spatial frequencies of the target, respectively.
In a traditional MIMO radar, since both the transmit and receive spatial frequencies are angle- but not range-dependent, the targets and interferences will be diagonally distributed in the joint transmit-receive domain, causing any mainlobe interferer to overlap with the response from the target such that these cannot be distinguished \cite{9161264}.
In contrast, the FDA transmit spatial frequency ${{\bf{u}}_{\mathop{\rm T}\nolimits} }\left( {{f_t} } \right)$ is range-angle-dependent, ensuring that any mainlobe interferer will be separated from the target in the joint transmit-receive domain, as long as such an interferer and the target are not at the same range.
As a result, this property can be exploited to suppress mainlobe interferers.

In order to do so, a beamformer or receive filter weight $\mathbf{w}\in {{\mathbb{C}}^{{{N}_{T}}{{N}_{R}}\tilde{L}\times 1}}$ is employed to synthesize the multiple outputs. Then, the output SINR ${{\kappa }_\text{FDA}}\left( \mathbf{s},\mathbf{w} \right)$ can be expressed as \cite{6649991}
\begin{equation}\label{eq:sinr}
\begin{aligned}
& {{\kappa }_{\text{FDA}}}\left( \mathbf{s},\mathbf{w} \right) \\ 
& =\frac{\mathsf{\mathbb{E}}\left\{ {{\left| \varsigma \left( {{r}_{t}},{{\theta }_{t}} \right){{\mathbf{w}}^{H}}\mathbf{\bar{A}}\left( {{f}_{t,t}},{{f}_{r,t}} \right)\mathbf{s} \right|}^{2}} \right\}}{\mathsf{\mathbb{E}}\left\{ {{\left| \sum\limits_{i=1}^{\Upsilon }{\varsigma \left( {{r}_{i}},{{\theta }_{i}} \right){{\mathbf{w}}^{H}}\mathbf{\bar{A}}\left( {{f}_{t,i}},{{f}_{r,i}} \right)\mathbf{s}} \right|}^{2}} \right\}+\mathsf{\mathbb{E}}\left\{ {{\left| {{\mathbf{w}}^{H}}\mathbf{n} \right|}^{2}} \right\}} \\ 
& =\frac{\text{SNR} \cdot {{\left| {{\mathbf{w}}^{H}}\mathbf{\bar{A}}\left( {{f}_{t,t}},{{f}_{r,t}} \right)\mathbf{s} \right|}^{2}}}{{{\mathbf{w}}^{H}}{{\mathbf{\Xi }}_{\Upsilon }}\left( \mathbf{s};{{f}_{t,i}},{{f}_{r,i}} \right)\mathbf{w}+{{\left\| \mathbf{w} \right\|}^{2}}}. \\ 
\end{aligned}
\end{equation}
where $\mathbf{\bar{A}}\left( {{f}_{t}},{{f}_{r}} \right)=\mathbf{K}{{\left( {{l}^{r}} \right)}^{T}}\otimes \left[ {{\mathbf{u}}_{\text{R}}}\left( {{f}_{r}} \right)\otimes \text{diag}\left\{ {{\mathbf{u}}_{\text{T}}}\left( {{f}_{t}},{{f}_{r}} \right) \right\} \right]$, ${{f}_{r,i}}=-\frac{{{d}_{r}}\sin {{\theta }_{i}}}{\lambda }$,
and
\begin{equation}
\begin{array}{*{35}{l}}
{{\mathbf{\Xi }}_{\Upsilon}}\left( \mathbf{s};{{f}_{t,i}},{{f}_{r,i}} \right)=  \\
\sum\limits_{i=1}^{\Upsilon}{\left\{ \text{IN}{{\text{R}}_{i}}\cdot \mathbf{\bar{A}}\left( {{f}_{t,i}},{{f}_{r,i}} \right)\mathbf{s}{{\mathbf{s}}^{H}}{{{\mathbf{\bar{A}}}}^{H}}\left( {{f}_{t,i}},{{f}_{r,i}} \right) \right\}}  \\
\end{array}
\end{equation}
where $\text{SNR}=\frac{\mathsf{\mathbb{E}}\left\{ {{\left| \varsigma \left( {{r}_{t}},{{\theta }_{t}} \right) \right|}^{2}} \right\}}{{{\sigma }^{2}}}$ and $\text{IN}{{\text{R}}_{i}}\text{=}\frac{\mathsf{\mathbb{E}}\left\{ {{\left| \varsigma \left( {{r}_{i}},{{\theta }_{i}} \right) \right|}^{2}} \right\}}{{{\sigma }^{2}}}$
represent the signal-to-noise ratio (SNR) and interference-to-noise ratio (INR) of the $i$th interference source at the input of the receiver, respectively, with ${{\sigma }^{2}}$ denoting the power of the additive noise.
The derivation in \eqref{eq:sinr} implicitly assumes that $\left\{ \varsigma \left( {{r}_{i}},{{\theta }_{i}} \right) \right\}_{i=1}^{I}$ are independent random variables. 
As a result, the mainlobe interference suppression problem can be formulated as
\begin{equation}
{{P}_{1}}\left\{ \begin{matrix}
\underset{\mathbf{w}}{\mathop{\max }}\, & \frac{{{\left| {{\mathbf{w}}^{H}}\mathbf{\bar{A}}\left( {{f}_{t,t}},{{f}_{r,t}} \right)\mathbf{s} \right|}^{2}}}{{{\mathbf{w}}^{H}}{{\mathbf{\Xi }}_{\Upsilon}}\left( \mathbf{s};{{f}_{t,i}},{{f}_{r,i}} \right)\mathbf{w}+{{\left\| \mathbf{w} \right\|}^{2}}}  \\
\end{matrix} \right..
\end{equation}
The problem $P_1$ is the well-known MVDR problem, with the optimal filter weights ${{\mathbf{w}}_{opt}}$ being given by \cite{6649991}
\begin{equation}\label{eq:weig}
{{\mathbf{w}}_\text{opt}}=\frac{\mathbf{Z}{{\left( \mathbf{s};{{f}_{t,i}},{{f}_{r,i}} \right)}^{-1}}\mathbf{\bar{A}}\left( {{f}_{t,t}},{{f}_{r,t}} \right)\mathbf{s}}{{{\mathbf{s}}^{H}}{{{\mathbf{\bar{A}}}}^{H}}\left( {{f}_{t,t}},{{f}_{r,t}} \right)\mathbf{Z}{{\left( \mathbf{s};{{f}_{t,i}},{{f}_{r,i}} \right)}^{-1}}\mathbf{\bar{A}}\left( {{f}_{t,t}},{{f}_{r,t}} \right)\mathbf{s}}
\end{equation}
with $\mathbf{Z}\left( \mathbf{s};{{f}_{t,i}},{{f}_{r,i}} \right)={{\mathbf{\Xi }}_{\Upsilon }}\left( \mathbf{s};{{f}_{t,i}},{{f}_{r,i}} \right)+{{\mathbf{I}}_{{{N}_{T}}{{N}_{R}}\tilde{L}}}$
representing the FDA interference-plus-noise covariance matrix.
Substituting ${{\mathbf{w}}_\text{opt}}$ into \eqref{eq:sinr}, some algebraic manipulations yields 
\begin{equation}
{{\kappa }_\text{FDA}}\left( \mathbf{s} \right)=\text{SNR}\cdot {{\mathbf{s}}^{H}}\mathbf{\Psi }\left( \mathbf{s} \right)\mathbf{s}
\end{equation}
where $\mathbf{\Psi }\left( \mathbf{s} \right)={{{\mathbf{\bar{A}}}}^{H}}\left( {{f}_{t,t}},{{f}_{r,t}} \right)\mathbf{Z}{{\left( \mathbf{s};{{f}_{t,i}},{{f}_{r,i}} \right)}^{-1}}\mathbf{\bar{A}}\left( {{f}_{t,t}},{{f}_{r,t}} \right).$

Note that the output SINR is therefore waveform-dependent. 
We proceed to considering how the FDA transmit waveform design can be incorporated in the formulation to further
improve output SINR.

\section{Problem formulation}\label{sec:b}
\subsection{Constraints}
\noindent 
The selection of a radar transmit waveform is constrained by various practical requirements, which directly affects the performance of the radar system, and is usually closely related to the mathematical characteristics of the actual optimization model.
Conventional constraints mainly include energy constraint \cite{1673421}, constant modulus \cite{6649991} or peak-to-average power ratio constraint \cite{8141978}, similarity constraint \cite{6649991}, and spectral constraint \cite{8356676}.
Considering the FDA transmission mechanism, such energy and similarity constraints are also imposed on the FDA transmit waveform.

\subsubsection{Energy constraint}
\noindent
Since the FDA essentially transmits a multi-carrier signal, the transmit energy of each transmit antenna must be constrained.
Without loss of generality, assuming that the energy of each transmit antenna is $\frac{1}{{{N}_{T}}}$, then the overall energy constraint can thus be written as \cite{8356676}
\begin{equation}
{{\int_{{{T}_{p}}}{\left| {{s}_{m}}\left( t \right){{e}^{-j2\pi \left( m-1 \right)\Delta ft}} \right|}}^{2}}\operatorname{d}t=\mathbf{S}\left( m \right){{\mathbf{S}}^{H}}\left( m \right)=\frac{1}{{{N}_{T}}}
\end{equation}
for $m=1,...,{{N}_{T}}$, where $\mathbf{S}\left( m \right)\in {{\mathbb{C}}^{L\times 1}}$ denotes the $m$th row of the transmit waveform matrix $\mathbf{S}$ and $T_p$ is pulse duration.
The above constraint can be alternatively written as
\begin{equation}
\mathbf{s}_{T}^{H}{{\mathbf{\Sigma }}_{m}}{{\mathbf{s}}_{T}}=\frac{1}{{{N}_{T}}}
\end{equation}
for $m=1,2,...,{{N}_{T}}$, where ${{\mathbf{\Sigma }}_{m}}\in {{\mathbb{C}}^{{{N}_{T}}L\times {{N}_{T}}L}}$
represents a block diagonal matrix, with its $m$th diagonal block being an identity matrix ${{\bf{I}}_{L}}$. Let
\begin{equation}
{{\bf{s}}_T} = {\mathop{\rm vec}\nolimits} \left\{ {{{\bf{S}}^T}} \right\} = {\bf{T}}\left( {{N_T},L} \right){\mathop{\rm vec}\nolimits} \left\{ {\bf{S}} \right\} = {\bf{T}}\left( {{N_T},L} \right){\bf{s}}
\end{equation}
with ${\bf{T}}\left( {{N_T},L} \right)$ being the commutation matrix for which the following properties hold \cite{greub2012linear}
\begin{subequations}
	\begin{equation}
	\kern 30pt {\bf{T}}\left( {{N_T},L} \right) = \sum\limits_{l = 1}^L {\left( {{\bf{e}}_l^T \otimes {{\bf{I}}_{{N_T}}} \otimes {{\bf{e}}_j}} \right)}
	\end{equation}
	\begin{equation}
	\kern -20pt {{\bf{T}}^T}\left( {{N_T},L} \right) = {\bf{T}}\left( {L,{N_T}} \right)
	\end{equation}
\end{subequations}

\subsubsection{Similarity constraint}
\noindent
Implementing similarity constraint on waveforms allows for a trade-off between obtaining optimal output and controlling other desired waveform properties such as pulse compression \cite{7450660,8141978}.
We consider the following similarity constraint
\begin{equation}\label{eq:simil}
{\left\| {{{\bf{s}}_T} - {{\bf{s}}_{{\mathop{\rm Ref}\nolimits} }}} \right\|_\infty } \le \varepsilon ,\kern 25pt \varepsilon  \ge 0.
\end{equation}
where ${{\bf{s}}_{{\mathop{\rm Ref}\nolimits} }}$ denotes the reference waveform and $\varepsilon$ is a user-defined parameter ruling the extent of the similarity.
Alternatively, \eqref{eq:simil} can be re-expressed as a series of quadratic inequality constraints given by
\begin{equation}
{{{\left( {{{\bf{s}}_T} - {{\bf{s}}_{{\mathop{\rm Ref}\nolimits} }}} \right)}^H}{{{\bf{\bar E}}}_j}\left( {{{\bf{s}}_T} - {{\bf{s}}_{{\mathop{\rm Ref}\nolimits} }}} \right) \le \varepsilon^2 ,\kern 5pt \varepsilon  \ge 0,}
\end{equation}
for $j = 1,...,NL$, where
\begin{equation}
\kern 0pt {{\mathbf{\bar{E}}}_{j}}\left( p,q \right)=\left\{ \begin{matrix}
1 & p=j,q=j  \\
0 & \text{otherwise}  \\
\end{matrix} \right..
\end{equation}

\subsubsection{Bandwidth constraint}
\noindent
The bandwidth of each transmit waveform is restricted to be less than the frequency increment of the FDA radar, being necessary due to the receiver being formed using the here proposed use of multi-carrier low pass filtering.
We define the energy spectral density (ESD) of the $m$th transmit waveform
\begin{equation}
\begin{array}{l}
{S_m}\left( f \right) = {\left| {\sum\limits_{l = 1}^L {{\bf{S}}\left( {m,l} \right){e^{ - j2\pi fl}}} } \right|^2}\\
\kern 32pt = {\left| {{{\bf{S}}^H}\left( m \right){{{\bf{\tilde e}}}_f}} \right|^2} = {\bf{S}}\left( m \right){{{\bf{\tilde e}}}_f}{\bf{\tilde e}}_f^H{{\bf{S}}^H}\left( m \right)\\
\end{array}
\end{equation}
for $m = 1,2,...,{N_T}$, where 
\begin{equation}
{{\bf{\tilde e}}_f} = {\left[ {{e^{ - j2\pi f}},{e^{ - j2\pi 2f}},...,{e^{ - j2\pi Lf}}} \right]^T}.
\end{equation}
Thus, the bandwidth constraint can be expressed as
\begin{equation}
\begin{array}{l}
\int_0^{{f_{m,lp}}} {{S_m}\left( f \right){\mathop{\rm d}\nolimits} f}  = \int_0^{f_{m,lp}} {{\bf{S}}\left( m \right){{\bf{\tilde e}}_f}{\bf{\tilde e}}_f^H{{\bf{S}}^H}\left( m \right){\mathop{\rm d}\nolimits} f}  \ge \frac{{{\gamma _m}}}{{{N_T}}},
\end{array}
\end{equation}
for $m = 1,2,...,{N_T}$, where $\int_0^{{f_{m,lp}}} {{S_m}\left( f \right){\mathop{\rm d}\nolimits} f}$ represents the energy of the $m$th transmit waveform after low pass filtering, and ${\gamma _m} \in \left( {0,1} \right]$ is a user-defined scalar that defines the tolerance for in-band energy, a typical choice being  ${\gamma _m} = 0.91$ \cite{mitra2006digital}.
Since
\begin{equation}
\begin{array}{l}
\kern 10pt {{\bf{H}}_m}\left( {p,q} \right) \in {\mathbb{C}^{L \times L}} = \int_0^{{f_{m,lp}}} {{{{\bf{\tilde e}}}_f}{\bf{\tilde e}}_f^H{\mathop{\rm d}\nolimits} f} \\
\kern 82pt = \left\{ {\begin{array}{*{20}{c}}
	{{f_{m,lp}}}&{p = q}\\
	{\frac{{{e^{j2\pi {f_{m,lp}}\left( {p - q} \right)}} - 1}}{{j2\pi \left( {p - q} \right)}}}&{p \ne q}
	\end{array}} \right.
\end{array}
\end{equation}
where ${{\bf{H}}_m}\left( {p,q} \right)$ denotes the $p$th row and $j$th column element of ${{\bf{H}}_m}$, the bandwidth constraint may equivalently be expressed as
\begin{equation}
{\bf{s}}_T^H{{\bf{B}}_m}{{\bf{s}}_T} \ge \frac{{{\gamma _m}}}{{{N_T}}}
\end{equation}
for $m = 1,2,...,{N_T}$, where ${{\mathbf{B}}_{m}}$
represents a block diagonal matrix, with its $m$th diagonal block being ${{\mathbf{H}}_{m}}$.

\subsection{Optimization problem}
\noindent
Based on the aforementioned discussions, the FDA transmit waveform design problem may be formulated as
\begin{equation}
{{{P}}_{2}}\left\{ \begin{matrix}
\underset{{{\mathbf{s}}_{T}}}{\mathop{\min }}\, & -\mathbf{s}_{T}^{H}\mathbf{\bar{\Psi }}\left( {{\mathbf{s}}_{T}} \right){{\mathbf{s}}_{T}}  \\
\operatorname{s}.t. & \left\{ \begin{matrix}
{{\left( {{\mathbf{s}}_{T}}-{{\mathbf{s}}_{\operatorname{Ref}}} \right)}^{H}}{{{\mathbf{\bar{E}}}}_{j}}\left( {{\mathbf{s}}_{T}}-{{\mathbf{s}}_{\operatorname{Ref}}} \right)\le \varepsilon^2  \\
\begin{aligned}
& \mathbf{s}_{T}^{H}{{\mathbf{\Sigma }}_{m}}{{\mathbf{s}}_{T}}=\frac{1}{{{N}_{T}}} \\ 
& \mathbf{s}_{T}^{H}{{\mathbf{B}}_{m}}{{\mathbf{s}}_{T}}\ge \frac{{{\gamma }_{m}}}{{{N}_{T}}} \\ 
\end{aligned}  \\
\end{matrix} \right.  \\
\end{matrix} \right.,
\end{equation}
for $j=1,2,...,{{N}_{T}}L$ and $m=1,2,...,{{N}_{T}}$, with $\varepsilon \ge 0$ and ${{\gamma }_{m}}\in \left( 0,1 \right]$, where 
\begin{subequations}
	\begin{equation}
	\kern 25pt \mathbf{\bar{\Psi }}\left( {{\mathbf{s}}_{T}} \right)={{\mathbf{\Lambda }}^{H}}\left( {{f}_{t}},{{f}_{r}} \right)\mathbf{\tilde{Z}}{{\left( {{\mathbf{s}}_{T}};{{f}_{t,i}},{{f}_{r,i}} \right)}^{-1}}\mathbf{\Lambda }\left( {{f}_{t}},{{f}_{r}} \right)
	\end{equation}
	\begin{equation}
	\begin{aligned}
	& \mathbf{\tilde{Z}}\left( {{\mathbf{s}}_{T}};{{f}_{t,i}},{{f}_{r,i}} \right)=\sum\limits_{i=1}^{\Upsilon}{\left\{ \text{IN}{{\text{R}}_{i}} \mathbf{\Lambda }\left( {{f}_{t}},{{f}_{r}} \right){{\mathbf{s}}_{T}}\mathbf{s}_{T}^{H}{{\mathbf{\Lambda }}^{H}}\left( {{f}_{t,i}},{{f}_{r,i}} \right) \right\}} \\ 
	& \kern 77pt +{{\mathbf{I}}_{{{N}_{T}}{{N}_{R}}\tilde{L}}} \\ 
	\end{aligned}
	\end{equation}
	\begin{equation}\label{eq:38c}
	\kern -62 pt \mathbf{\Lambda }\left( {{f}_{t}},{{f}_{r}} \right)=\mathbf{\bar{A}}\left( {{f}_{t}},{{f}_{r}} \right)\mathbf{T}\left( L,{{N}_{T}} \right)
	\end{equation}
\end{subequations}

It can be seen that problem $P_2$, containing a non-convex objective function, multiple nonlinear equality (the energy constraint) and non-convex quadratic inequality constraints (the bandwidth constraints), is non-convex and can therefore not be solved in polynomial time \cite{boyd2004convex}.
In the traditional ADMM algorithm structure, the update of the primary and auxiliary variables needs to ensure that the optimal solution is produced \cite{boyd2011distributed}.
However, the complexity of the objective function is the main factor that makes the problem $P_2$ difficult to solve with the traditional ADMM algorithm.
Here, we employ two tricks resulting in the P-ADMM and MM-ADMM algorithms.
We do not directly optimize the objective function but instead optimizes a simple surrogate function.
At each iteration of the P-ADMM algorithm, the previous iteration results are used in each iteration to simplify the objective function. 
Inspired by the majorization-minimization (MM) method, in the MM-ADMM algorithm, the selected surrogate function locally approximates the objective function with their difference being minimized at the current point. 

\section{Optimization algorithm}\label{sec:c}
\subsection{P-ADMM}
\noindent
An auxiliary variable $\mathbf{h}$ is first introduced to reformulate $P_2$ as
\begin{equation}
{{{P}}_{3}}\left\{ \begin{matrix}
\underset{{{\mathbf{s}}_{T}}}{\mathop{\min }}\, & -\mathbf{s}_{T}^{H}\mathbf{\bar{\Psi }}\left( {{\mathbf{s}}_{T}} \right)\mathbf{h}  \\
\operatorname{s}.t. & \left\{ \begin{matrix}
\begin{aligned}
& \kern 33pt {{\mathbf{s}}_{T}}=\mathbf{h} \\ 
& {{\left( {{\mathbf{s}}_{T}}-{{\mathbf{s}}_{\operatorname{Ref}}} \right)}^{H}}{{{\mathbf{\bar{E}}}}_{j}}\left( \mathbf{h}-{{\mathbf{s}}_{\operatorname{Ref}}} \right)\le \varepsilon^2 \\ 
\end{aligned}  \\
\begin{aligned}
& \mathbf{s}_{T}^{H}{{\mathbf{\Sigma }}_{m}}\mathbf{h}=\frac{1}{{{N}_{T}}} \\ 
& \kern 1pt \mathbf{s}_{T}^{H}{{\mathbf{B}}_{m}}\mathbf{h}\ge \frac{{{\gamma }_{m}}}{{{N}_{T}}}\\ 
\end{aligned}  \\
\end{matrix} \right.  \\
\end{matrix} \right.
\end{equation}
It is worth noting that the non-convex energy and the bandwidth constraints on ${\mathbf{s}_T}$ in problem $P_2$ are transformed into affine constraints on ${\mathbf{s}_T}$ and $\mathbf{h}$ in problem $P_3$. Then, the scaled augmented Lagrangian function \cite{boyd2011distributed} of problem $P_3$ is formed as
\begin{equation}
\begin{aligned}
& {{\mathsf{\mathbb{L}}}_\text{P-ADMM}}\left( {{\mathbf{s}}_{T}},\mathbf{h};\mathbf{u},\left\{ {{\mathbf{v}}_{m}} \right\}_{m=1}^{{{N}_{T}}} \right) \\ 
& \kern 10pt =-\mathbf{s}_{T}^{H}\mathbf{\bar{\Psi }}\left( {{\mathbf{s}}_{T}} \right)\mathbf{h}+\frac{{{\rho }_{1}}}{2}\left\| {{\mathbf{s}}_{T}}-\mathbf{h}+\mathbf{u} \right\|_{2}^{2} \\ 
& \kern 20pt +\sum\limits_{m=1}^{{{N}_{T}}}{\left\{ \frac{{{\rho }_{2}}}{2}\left\| \mathbf{s}_{T}^{H}{{\mathbf{\Sigma }}_{m}}\mathbf{h}-\frac{1}{{{N}_{T}}}+{{\mathbf{v}}_{m}} \right\|_{2}^{2} \right\}} \\ 
\end{aligned}
\end{equation}
where $\mathbf{u}$ and $\left\{ {{\mathbf{v}}_{m}} \right\}_{m=1}^{{{N}_{T}}}$ are dual variables and Lagrange multipliers associated with the equality constraints ${{\mathbf{s}}_{T}}=\mathbf{h}$ and $\mathbf{s}_{T}^{H}{{\mathbf{\Sigma }}_{m}}\mathbf{h}=\frac{1}{{{N}_{T}}},m=1,2,...,{{N}_{T}}$, and ${{\rho }_{1}}$ and ${{\rho }_{2}}$ are penalty parameters.
Thus, the use of a large penalty parameter values imposes a large penalty on violations of the primal feasibility and will thus tend to yield small primal residuals and vice versa.
Technically, the ADMM blends the decomposability of the dual ascent with the superior convergence properties of the method of multipliers, with the primal and dual variables being updated in an sequential fashion by the use of augmented Lagrangian methods \cite{boyd2011distributed}.

The resulting update procedure of the P-ADMM algorithm is slightly different from the standard ADMM formulation.
At the $k+1$th iteration, P-ADMM consists of the following update procedures:
\begin{subequations}
	\begin{equation}\label{eq:pada}
 \mathbf{s}_{T}^{k+1}:=\underset{D\left( {{\mathbf{s}}_{T}} \right)}{\mathop{\arg \min }}\,\left\{ \begin{aligned}
	& -\mathbf{s}_{T}^{H}\mathbf{\bar{\Psi }}\left( \mathbf{s}_{T}^{k} \right){{\mathbf{h}}^{k}}+\frac{{{\rho }_{1}}}{2}\left\| {{\mathbf{s}}_{T}}-{{\mathbf{h}}^{k}}+{{\mathbf{u}}^{k}} \right\|_{2}^{2} \\ 
	& +\sum\limits_{m=1}^{{{N}_{T}}}{\left\{ \frac{{{\rho }_{2}}}{2}\left\| \mathbf{s}_{T}^{H}{{\mathbf{\Sigma }}_{m}}{{\mathbf{h}}^{k}}-\frac{1}{{{N}_{T}}}+\mathbf{v}_{m}^{k} \right\|_{2}^{2} \right\}} \\ 
	\end{aligned} \right.
	\end{equation}
	\begin{equation}\label{eq:padb}
	\kern -36pt {{\mathbf{h}}^{k+1}}:=\underset{D\left( \mathbf{h} \right)}{\mathop{\arg \min }}\,{{\mathsf{\mathbb{L}}}_\text{P-ADMM}}\left( \mathbf{s}_{T}^{k+1},\mathbf{h};{{\mathbf{u}}^{k}},\left\{ \mathbf{v}_{m}^{k} \right\}_{m=1}^{{{N}_{T}}} \right)
	\end{equation}
	\begin{equation}
	\kern -122pt {{\mathbf{u}}^{k+1}}:={{\mathbf{u}}^{k+1}}+\left( \mathbf{s}_{T}^{k+1}-{{\mathbf{h}}^{k+1}} \right)
	\end{equation}
	\begin{equation}
	\kern -70pt \mathbf{v}_{m}^{k+1}:=\mathbf{v}_{m}^{k+1}+\left[ {{\left( \mathbf{s}_{T}^{H} \right)}^{k+1}}{{\mathbf{\Sigma }}_{m}}{{\mathbf{h}}^{k+1}}-\frac{1}{{{N}_{T}}} \right]
	\end{equation}
\end{subequations}
for $m=1,2,...,{{N}_{T}}$, where the sets $D\left( {{\mathbf{s}}_{T}} \right)$ and $D\left( \mathbf{h} \right)$ are defined as
\begin{subequations}
	\begin{equation}
	D\left( {{\mathbf{s}}_{T}} \right)=\left\{ {{\mathbf{s}}_{T}}\left| \begin{matrix}
	{{\left( {{\mathbf{s}}_{T}}-{{\mathbf{s}}_{\operatorname{Ref}}} \right)}^{H}}{{{\mathbf{\bar{E}}}}_{j}}\left( \mathbf{h}-{{\mathbf{s}}_{\operatorname{Ref}}} \right)\le \varepsilon^2 \\
	\begin{aligned}
	& \mathbf{s}_{T}^{H}{{\mathbf{B}}_{m}}\mathbf{h}\ge \frac{{{\gamma }_{m}}}{{{N}_{T}}}\\ 
	\end{aligned}  \\
	\end{matrix} \right. \right\}
	\end{equation}
	\begin{equation}
	D\left( \mathbf{h} \right)=\left\{ \mathbf{h}\left| \begin{matrix}
	{{\left( {{\mathbf{s}}_{T}}-{{\mathbf{s}}_{\operatorname{Ref}}} \right)}^{H}}{{{\mathbf{\bar{E}}}}_{j}}\left( \mathbf{h}-{{\mathbf{s}}_{\operatorname{Ref}}} \right)\le \varepsilon^2 \\
	\begin{aligned}
	& \mathbf{s}_{T}^{H}{{\mathbf{B}}_{m}}\mathbf{h}\ge \frac{{{\gamma }_{m}}}{{{N}_{T}}} \\ 
	\end{aligned}  \\
	\end{matrix} \right. \right\}
	\end{equation}
\end{subequations}
It is worth noting that both the ${{\mathbf{s}}_{T}}$-update in \eqref{eq:pada} and the $\mathbf{h}$-update in \eqref{eq:padb}, which involve solving quadratic programming (QP) problems that contain multiple linear constraints, do not appear to admit a closed-form solution.
However, both problems may be solved efficiently by using the active set method (ASM) \cite{nocedal2006numerical} or CVX in MATLAB \cite{grant2014cvx}.

The proposed P-ADMM algorithm is summarized as follows:
\begin{breakablealgorithm} 
	\renewcommand{\algorithmicrequire}{\textbf{Input:}}
	\renewcommand{\algorithmicensure}{\textbf{Output:}}
	\caption{The P-ADMM algorithm.}
	\begin{algorithmic}[1] 
		\REQUIRE For $k = 0$, set the initial variables $\mathbf{s}_{T}^{0},{{\mathbf{h}}^{0}};{{\mathbf{u}}^{0}},\left\{ \mathbf{v}_{m}^{0} \right\}_{m=1}^{{{N}_{T}}}$ and the penalty parameters ${{\rho }_{1}}$ and ${{\rho }_{2}}$.\\ 
		\ENSURE The optimal transmit waveforms and receive filter weight pair $\left( {{\mathbf{s}}_{T,opt}},{{\mathbf{w}}_{opt}} \right)$.
		\STATE Let $k = k+1$ and update $\mathbf{s}_{T}^{k}$ by solving problem \eqref{eq:pada}.
		\STATE Update $\mathbf{h}^{k}$ by solving problem \eqref{eq:padb}.
		\STATE Compute \\ 
		$\kern 45pt {{\mathbf{u}}^{k}}={{\mathbf{u}}^{k-1}}+\left( \mathbf{s}_{T}^{k}-{{\mathbf{h}}^{k-1}} \right)$.
		\STATE Compute \\ 
		$\kern 7pt \mathbf{v}_{m}^{k}=\mathbf{v}_{m}^{k-1}+\left[ {{\left( \mathbf{s}_{T}^{H} \right)}^{k}}{{\mathbf{\Sigma }}_{m}}{{\mathbf{h}}^{k}}-\frac{1}{{{N}_{T}}} \right],m=1,2,...,{{N}_{T}}$.
		\STATE If the termination conditions are satified, output ${{\mathbf{s}}_\text{T,opt}}=\mathbf{s}_{T}^{k}$ and substitute ${{\mathbf{s}}_\text{T,opt}}$ into \eqref{eq:weig} to obtain ${{\mathbf{w}}_\text{opt}}$. Otherwise, repeat Step 1.
	\end{algorithmic}  
\end{breakablealgorithm}

\subsubsection{Computational complexity and convergence}
\noindent
Since the overall complexity of the P-ADMM algorithm is linear with the number of iterations, we here specify the complexity of each iteration.
Updating $\mathbf{s}_{T}^{k+1}$ or ${{\mathbf{h}}^{k}}$ by solving a quadratic programming (QP) problem with ${{N}_{T}}+{{N}_{T}}L$ inequality constraints may be implemented using the interior point method (IPM) in the CVX toolbox, 
The number of iterations is $\mathcal{O}\left( \sqrt{{{N}_{T}}+{{N}_{T}}L}\log \left( {1}/{\upsilon }\; \right) \right)$, where $\upsilon $ is the convergence parameter. In each iteration of the IPM, the computational complexity is $\mathcal{O}\left( N_{T}^{3.5}{{L}^{3.5}} \right)$.
Overall, in each iteration of the P-ADMM, the complexity is $\mathcal{O}\left( {{I}_{1}}\sqrt{{{N}_{T}}+{{N}_{T}}L}\log \left( {1}/{\upsilon }\; \right)\left( N_{T}^{3.5}{{L}^{3.5}} \right) \right)$, where ${{I}_{1}}$ denotes number of iterations of the P-ADMM.
For MIMO radar waveform design with signal-dependent non-mainlobe interference, the feasible solution set partition-based successive QCQP refinement-binary search
(SQR-BS) and non-decreasing (SQR-ND) algorithms proposed in \cite{7450660} are the most efficient.
The computational complexity of the algorithm is
$\mathcal{O}\left(F\left( N_{T}^{3.5}{{L}^{3.5}} \right) \right)$ with $F$ denoting the refinement steps, which is smaller than that of the SDR method even with $F=4$.
Although the convergence of the P-ADMM algorithm cannot be guaranteed, numerical simulations demonstrate the performance of the algorithm.

\subsection{MM-ADMM}
\noindent
Follow the results of \cite{8239836}, a suitable majorizer of the objective function $-\mathbf{s}_{T}^{H}\mathbf{\bar{\Psi }}\left( {{\mathbf{s}}_{T}} \right){{\mathbf{s}}_{T}}$ at point ${{{\mathbf{\tilde{s}}}}_{T}}$ can be expressed as 
\begin{equation}
\begin{aligned}
& M\left( {{\mathbf{s}}_{T}},{{{\mathbf{\tilde{s}}}}_{T}};{{\mathbf{S}}_{T}},{{{\mathbf{\tilde{S}}}}_{T}} \right) \\ 
& =-2\operatorname{Re}\left\{ {{\mathbf{z}}^{H}}\left( {{{\mathbf{\tilde{s}}}}_{T}};{{{\mathbf{\tilde{S}}}}_{T}} \right){{\mathbf{s}}_{T}} \right\}+2\operatorname{Tr}\left\{ \mathbf{P}\left( {{{\mathbf{\tilde{S}}}}_{T}} \right){{\mathbf{S}}_{T}} \right\} \\ 
& \kern 12pt -2\operatorname{Tr}\left\{ \mathbf{P}\left( {{{\mathbf{\tilde{S}}}}_{T}} \right){{{\mathbf{\tilde{S}}}}_{T}} \right\}+\mathbf{\tilde{s}}_{T}^{H}\mathbf{\bar{\Psi }}\left( {{{\mathbf{\tilde{s}}}}_{T}} \right){{{\mathbf{\tilde{s}}}}_{T}} \\ 
\end{aligned}
\end{equation}
where
\begin{subequations}
	\begin{equation}
	\kern 23pt \mathbf{z}\left( {{{\mathbf{\tilde{s}}}}_{T}};{{{\mathbf{\tilde{S}}}}_{T}} \right)={{\mathbf{\Lambda }}^{H}}\left( {{f}_{t,t}},{{f}_{r,t}} \right)\mathbf{\tilde{Z}}{{\left( {{{\mathbf{\tilde{s}}}}_{T}};{{f}_{t,i}},{{f}_{r,i}} \right)}^{-1}}\mathbf{\Lambda }\left( {{f}_{t,t}},{{f}_{r,t}} \right){{{\mathbf{\tilde{s}}}}_{T}}
	\end{equation}
	\begin{equation}
	\kern -44pt \mathbf{P}\left( {{{\mathbf{\tilde{S}}}}_{T}} \right)=\sum\limits_{i=1}^{I}{\text{IN}{{\text{R}}_{i}}\mathbf{Q}_{i}^{H}\left( {{{\mathbf{\tilde{S}}}}_{T}} \right){{{\mathbf{\tilde{S}}}}_{T}}{{\mathbf{Q}}_{i}}\left( {{{\mathbf{\tilde{S}}}}_{T}} \right)},
	\end{equation}
	\begin{equation}
	\kern 10pt {{\mathbf{Q}}_{i}}\left( {{{\mathbf{\tilde{S}}}}_{T}} \right)={{\mathbf{\Lambda }}^{H}}\left( {{f}_{t,t}},{{f}_{r,t}} \right)\mathbf{\tilde{Z}}{{\left( {{{\mathbf{\tilde{S}}}}_{T}};{{f}_{t,i}},{{f}_{r,i}} \right)}^{-1}}\mathbf{\Lambda }\left( {{f}_{t,t}},{{f}_{r,t}} \right).
	\end{equation}
\end{subequations}
and ${{{\mathbf{\tilde{S}}}}_{T}}={{{\mathbf{\tilde{s}}}}_{T}}\mathbf{\tilde{s}}_{T}^{H}$.
Therefore, the problem $P_2$ can be approximately expressed as
\begin{equation}
{{{P}}_{4}}\left\{ \begin{matrix}
\underset{{{\mathbf{s}}_{T}}}{\mathop{\min }}\, & \left\{ \begin{aligned}
& M\left( {{\mathbf{s}}_{T}},{{{\mathbf{\tilde{s}}}}_{T}};{{\mathbf{S}}_{T}},{{{\mathbf{\tilde{S}}}}_{T}} \right)\overset{\arg \min }{\mathop{=}}\,2{{\mathbf{s}}_{T}}\mathbf{P}\left( {{{\mathbf{\tilde{S}}}}_{T}} \right)\mathbf{s}_{T}^{H} \\ 
& \kern 63pt -2\operatorname{Re}\left\{ {{\mathbf{z}}^{H}}\left( {{{\mathbf{\tilde{s}}}}_{T}};{{{\mathbf{\tilde{S}}}}_{T}} \right){{\mathbf{s}}_{T}} \right\} \\ 
\end{aligned} \right.  \\
\operatorname{s}.t. & \left\{ \begin{matrix}
{{\left( {{\mathbf{s}}_{T}}-{{\mathbf{s}}_{\operatorname{Ref}}} \right)}^{H}}{{{\mathbf{\bar{E}}}}_{j}}\left( {{\mathbf{s}}_{T}}-{{\mathbf{s}}_{\operatorname{Ref}}} \right)\le \varepsilon^2  \\
\begin{aligned}
& \mathbf{s}_{T}^{H}{{\mathbf{\Sigma }}_{m}}{{\mathbf{s}}_{T}}=\frac{1}{{{N}_{T}}} \\ 
& \mathbf{s}_{T}^{H}{{\mathbf{B}}_{m}}{{\mathbf{s}}_{T}}\ge \frac{{{\gamma }_{m}}}{{{N}_{T}}} \\ 
\end{aligned}  \\
\end{matrix} \right.  \\
\end{matrix} \right.
\end{equation}
for $j=1,2,...,{{N}_{T}}L$ and $m=1,2,...,{{N}_{T}}$, with $\varepsilon \ge 0$ and ${{\gamma }_{m}}\in \left( 0,1 \right]$.
At the $(k+1)$th iteration, the update procedures of the MM-ADMM are given by
\begin{subequations}
	\begin{equation}\label{49a}
\kern -20pt \mathbf{s}_{T,r}^{k+1}:=\arg \min \left\{ \begin{aligned}
& 2{{\mathbf{s}}_{T,r}}{{\mathbf{P}}_{r}}\left( \mathbf{S}_{T}^{k} \right)\mathbf{s}_{T,r}^{H}-2\mathbf{z}_{r}^{T}\left( \mathbf{s}_{T}^{k};\mathbf{S}_{T}^{k} \right){{\mathbf{s}}_{T,r}} \\ 
& +\sum\limits_{j=1}^{{{N}_{T}}L}{\left\{ \frac{{{\rho }_{1}}}{2}\left\| {{{\mathbf{\bar{E}}}}_{j,r}}\left( {{\mathbf{s}}_{T,r}}-{{\mathbf{s}}_{\operatorname{Ref},r}} \right)-\mathbf{h}_{j,r}^{k}+\mathbf{p}_{j,r}^{k} \right\|_{2}^{2} \right\}} \\ 
& +\sum\limits_{m=1}^{{{N}_{T}}}{\left\{ \frac{{{\rho }_{2}}}{2}\left\| {{\mathbf{\Sigma }}_{m,r}}{{\mathbf{s}}_{T,r}}-\mathbf{u}_{m,r}^{k}+\mathbf{q}_{m,r}^{k} \right\|_{2}^{2} \right\}} \\ 
& +\sum\limits_{m=1}^{{{N}_{T}}}{\left\{ \frac{{{\rho }_{3}}}{2}\left\| {{{\mathbf{\bar{B}}}}_{m,r}}{{\mathbf{s}}_{T,r}}-\mathbf{v}_{m,r}^{k}+\mathbf{d}_{m,r}^{k} \right\|_{2}^{2} \right\}} \\ 
\end{aligned} \right\}
	\end{equation}
	\begin{equation}\label{49b}
	\mathbf{h}_{j,r}^{k+1}:=\underset{\left\| {{\mathbf{h}}_{j}} \right\|_{2}^{2}\le {{\varepsilon }^{2}},\varepsilon \ge 0}{\mathop{\arg \min }}\,{{\mathsf{\mathbb{L}}}_{\text{MM-ADMM}}}\left( \begin{aligned}
	& \mathbf{s}_{T,r}^{k+1},\left\{ {{\mathbf{h}}_{j,r}} \right\}_{j=1}^{{{N}_{T}}L},\left\{ \mathbf{u}_{m,r}^{k} \right\}_{m=1}^{{{N}_{T}}},\left\{ \mathbf{v}_{m,r}^{k} \right\}_{m=1}^{{{N}_{T}}}; \\ 
	& \left\{ \mathbf{p}_{j,r}^{k} \right\}_{j=1}^{{{N}_{T}}L},\left\{ \mathbf{q}_{m,r}^{k} \right\}_{m=1}^{{{N}_{T}}},\left\{ \mathbf{d}_{m,r}^{k} \right\}_{m=1}^{{{N}_{T}}} \\ 
	\end{aligned} \right)
	\end{equation}
	\begin{equation}
\mathbf{u}_{m,r}^{k+1}:=\underset{\left\| {{\mathbf{u}}_{m}} \right\|_{2}^{2}=\frac{1}{{{N}_{T}}}}{\mathop{\arg \min }}\,{{\mathsf{\mathbb{L}}}_{\text{MM-ADMM}}}\left( \begin{aligned}
& \mathbf{s}_{T,r}^{k+1},\left\{ \mathbf{h}_{j,r}^{k+1} \right\}_{j=1}^{{{N}_{T}}L},\left\{ {{\mathbf{u}}_{m,r}} \right\}_{m=1}^{{{N}_{T}}},\left\{ \mathbf{v}_{m,r}^{k} \right\}_{m=1}^{{{N}_{T}}}; \\ 
& \left\{ \mathbf{p}_{j,r}^{k} \right\}_{j=1}^{{{N}_{T}}L},\left\{ \mathbf{q}_{m,r}^{k} \right\}_{m=1}^{{{N}_{T}}},\left\{ \mathbf{d}_{m,r}^{k} \right\}_{m=1}^{{{N}_{T}}} \\ 
\end{aligned} \right)
	\end{equation}
	\begin{equation}
 \mathbf{v}_{m,r}^{k+1}:=\underset{\begin{matrix}
		\left\| {{\mathbf{v}}_{m}} \right\|_{2}^{2}\ge \frac{{{\gamma }_{m}}}{{{N}_{T}}}  \\
		{{\gamma }_{m}}\in \left( 0,1 \right]  \\
		\end{matrix}}{\mathop{\arg \min }}\,{{\mathsf{\mathbb{L}}}_{\text{MM-ADMM}}}\left( \begin{aligned}
	& \mathbf{s}_{T,r}^{k+1},\left\{ \mathbf{h}_{j,r}^{k+1} \right\}_{j=1}^{{{N}_{T}}L},\left\{ \mathbf{u}_{m,r}^{k+1} \right\}_{m=1}^{{{N}_{T}}},\left\{ {{\mathbf{v}}_{m,r}} \right\}_{m=1}^{{{N}_{T}}}; \\ 
	& \left\{ \mathbf{p}_{j,r}^{k} \right\}_{j=1}^{{{N}_{T}}L},\left\{ \mathbf{q}_{m,r}^{k} \right\}_{m=1}^{{{N}_{T}}},\left\{ \mathbf{d}_{m,r}^{k} \right\}_{m=1}^{{{N}_{T}}} \\ 
	\end{aligned} \right)
	\end{equation}
	\begin{equation} \kern -70pt \mathbf{p}_{j,r}^{k+1}:=\mathbf{p}_{j,r}^{k}+\left[ {{{\mathbf{\bar{E}}}}_{j,r}}\left( \mathbf{s}_{T,r}^{k+1}-{{\mathbf{s}}_{\operatorname{Ref},r}} \right)-\mathbf{h}_{j,r}^{k+1} \right],j=1,2,...,{{N}_{T}}L
	\end{equation}
	\begin{equation} \kern -110pt \mathbf{q}_{m,r}^{k+1}:=\mathbf{q}_{m,r}^{k}+\left( {{\mathbf{\Sigma }}_{m,r}}\mathbf{s}_{T,r}^{k+1}-\mathbf{u}_{m,r}^{k+1} \right),m=1,2,...,{{N}_{T}}
	\end{equation}
	\begin{equation} \kern -110pt \mathbf{d}_{m,r}^{k+1}:=\mathbf{d}_{m,r}^{k}+\left( {{{\mathbf{\bar{B}}}}_{m,r}}\mathbf{s}_{T,r}^{k+1}-\mathbf{v}_{m,r}^{k+1} \right),m=1,2,...,{{N}_{T}}
	\end{equation}
\end{subequations}

\emph{Proof}: See \ref{sec:appenMMx}.
\subsubsection{Update ${{\mathbf{s}}_{T,r}}$}
\noindent
Let $G\left( {{\mathbf{s}}_{T,r}} \right)$ denote the objective function of \eqref{49a}. 
Then, the gradient of $G\left( {{\mathbf{s}}_{T,r}} \right)$ is given by
\begin{equation}
\begin{aligned}
& {{\nabla }_{{{\mathbf{s}}_{T,r}}}}G\left( {{\mathbf{s}}_{T,r}} \right)= \\ 
& 2\left[ \mathbf{P}_{r}^{T}\left( \mathbf{S}_{T}^{k} \right)+{{\mathbf{P}}_{r}}\left( \mathbf{S}_{T}^{k} \right) \right]{{\mathbf{s}}_{T,r}}-2{{\mathbf{z}}_{r}}\left( \mathbf{s}_{T}^{k};\mathbf{S}_{T}^{k} \right) \\ 
& +{{\rho }_{1}}\sum\limits_{j=1}^{{{N}_{T}}L}{\left\{ \mathbf{\bar{E}}_{j,r}^{T}\left( {{{\mathbf{\bar{E}}}}_{j,r}}\left( {{\mathbf{s}}_{T,r}}-{{\mathbf{s}}_{\operatorname{Ref},r}} \right)-\mathbf{h}_{j,r}^{k}+\mathbf{p}_{j,r}^{k} \right) \right\}} \\ 
& +{{\rho }_{2}}\sum\limits_{m=1}^{{{N}_{T}}}{\left\{ \mathbf{\Sigma }_{m,r}^{T}\left( {{\mathbf{\Sigma }}_{m,r}}{{\mathbf{s}}_{T,r}}-\mathbf{u}_{m,r}^{k}+\mathbf{q}_{m,r}^{k} \right) \right\}} \\ 
& +{{\rho }_{3}}\sum\limits_{m=1}^{{{N}_{T}}}{\left\{ \mathbf{\bar{B}}_{m,r}^{T}\left( {{{\mathbf{\bar{B}}}}_{m,r}}{{\mathbf{s}}_{T,r}}-\mathbf{v}_{m,r}^{k}+\mathbf{d}_{m,r}^{k} \right) \right\}} \\ 
\end{aligned}
\end{equation}
From the first order optimality conditions \cite{boyd2004convex}, we obtain the $\mathbf{s}_{T}^{k+1}$ update as
	\begin{equation}\label{eq:51}
\begin{aligned}
& \mathbf{s}_{T}^{k+1}=\frac{1}{2}{{\left[ \mathbf{P}_{r}^{T}\left( \mathbf{S}_{T}^{k} \right)+{{\mathbf{P}}_{r}}\left( \mathbf{S}_{T}^{k} \right)+{{\rho }_{1}}\sum\limits_{j=1}^{{{N}_{T}}L}{\left\{ {{{\mathbf{\bar{E}}}}_{j,r}} \right\}}+{{\rho }_{2}}\sum\limits_{m=1}^{{{N}_{T}}}{\left\{ {{\mathbf{\Sigma }}_{m,r}} \right\}}+{{\rho }_{3}}\sum\limits_{m=1}^{{{N}_{T}}}{{{\mathbf{B}}_{m,r}}} \right]}^{-1}} \\ 
& \kern 25pt \cdot \times \left( \begin{aligned}
& 2{{\mathbf{z}}_{r}}\left( \mathbf{s}_{T}^{k};\mathbf{S}_{T}^{k} \right)+{{\rho }_{1}}\sum\limits_{j=1}^{{{N}_{T}}L}{\left\{ \mathbf{\bar{E}}_{j,r}^{T}\left( {{{\mathbf{\bar{E}}}}_{j,r}}{{\mathbf{s}}_{\operatorname{Ref}}}+\mathbf{h}_{j,r}^{k}-\mathbf{p}_{j,r}^{k} \right) \right\}} \\ 
& +{{\rho }_{2}}\sum\limits_{m=1}^{{{N}_{T}}}{\left\{ \mathbf{\Sigma }_{m,r}^{T}\left( \mathbf{u}_{m,r}^{k}-\mathbf{q}_{m,r}^{k} \right) \right\}}+{{\rho }_{3}}\sum\limits_{m=1}^{{{N}_{T}}}{\left\{ \mathbf{\bar{B}}_{m,r}^{T}\left( \mathbf{v}_{m,r}^{k}-\mathbf{d}_{m,r}^{k} \right) \right\}} \\ 
\end{aligned} \right) \\ 
\end{aligned}
\end{equation}

\subsubsection{Update ${{\mathbf{h}}_{j,r}}$, ${{\mathbf{u}}_{m,r}}$, and ${{\mathbf{v}}_{m,r}}$}
\noindent
The update process for different auxiliary variables is similar, with their updates being computed as
\begin{subequations}
	\begin{equation}\label{eq:52a}
	\kern -5pt \mathbf{h}_{j,r}^{k+1}=\left\{ \begin{matrix}
	{{h}_{j}}\left( \mathbf{s},\mathbf{p} \right) & \left\| {{h}_{j}}\left( \mathbf{s},\mathbf{p} \right) \right\|_{2}^{2}\le \varepsilon^2   \\
	\frac{{{h}_{j}}\left( \mathbf{s},\mathbf{p} \right){\varepsilon }}{\left\| {{h}_{j}}\left( \mathbf{s},\mathbf{p} \right) \right\|_{2}^{2}} & \left\| {{h}_{j}}\left( \mathbf{s},\mathbf{p} \right) \right\|_{2}^{2}\ge \varepsilon^2   \\
	\end{matrix} \right.
	\end{equation}
	\begin{equation}\label{eq:52b}
	\kern 25pt \mathbf{u}_{m,r}^{k+1}=\left\{ \begin{matrix}
	{{u}_{m}}\left( \mathbf{s},\mathbf{q} \right) & \left\| {{u}_{m}}\left( \mathbf{s},\mathbf{q} \right) \right\|_{2}^{2}=\frac{1}{{{N}_{T}}}  \\
	\frac{{{u}_{m}}\left( \mathbf{s},\mathbf{q} \right)}{\left\| {{u}_{m}}\left( \mathbf{s},\mathbf{q} \right) \right\|_{2}^{2}\sqrt{{{N}_{T}}}} & \left\| {{u}_{m}}\left( \mathbf{s},\mathbf{q} \right) \right\|_{2}^{2}\ne \frac{1}{{{N}_{T}}}  \\
	\end{matrix} \right.
	\end{equation}
	\begin{equation}\label{eq:52c}
	\kern 25pt \mathbf{v}_{m,r}^{k+1}=\left\{ \begin{matrix}
	{{v}_{m}}\left( \mathbf{s},\mathbf{d} \right) & \left\| {{v}_{m}}\left( \mathbf{s},\mathbf{d} \right) \right\|_{2}^{2}\ge \frac{{{\gamma }_{m}}}{{{N}_{T}}}  \\
	\frac{{{v}_{m}}\left( \mathbf{s},\mathbf{d} \right)\sqrt{{{\gamma }_{m}}}}{\left\| {{v}_{m}}\left( \mathbf{s},\mathbf{d} \right) \right\|_{2}^{2}\sqrt{{{N}_{T}}}} & \left\| {{v}_{m}}\left( \mathbf{s},\mathbf{d} \right) \right\|_{2}^{2}\le \frac{{{\gamma }_{m}}}{{{N}_{T}}}  \\
	\end{matrix} \right.
	\end{equation}
\end{subequations}
where
\begin{subequations}
	\begin{equation}
\kern 18pt	h\left( \mathbf{s}_{T,r}^{k+1},\mathbf{p}_{j,r}^{k} \right)={{{\mathbf{\bar{E}}}}_{j,r}}\left( \mathbf{s}_{T,r}^{k+1}-{{\mathbf{s}}_{\operatorname{Ref},r}} \right)+\mathbf{p}_{j,r}^{k}
	\end{equation}
	\begin{equation}
	{{u}_{m}}\left( \mathbf{s},\mathbf{q} \right)={{\mathbf{\Sigma }}_{m,r}}\mathbf{s}_{T,r}^{k+1}+\mathbf{q}_{m,r}^{k}
	\end{equation}
	\begin{equation}
	{{v}_{m}}\left( \mathbf{s},\mathbf{d} \right)={{{\mathbf{\bar{B}}}}_{m,r}}\mathbf{s}_{m,r}^{k+1},+\mathbf{d}_{m,r}^{k}
	\end{equation}
\end{subequations}

\emph{Proof}: See \ref{sec:appen}.

The proposed MM-ADMM algorithm is summarized as follows:
\begin{breakablealgorithm} 
	\renewcommand{\algorithmicrequire}{\textbf{Input:}}
	\renewcommand{\algorithmicensure}{\textbf{Output:}}
	\caption{The MM-ADMM algorithm.}
	\begin{algorithmic}[1] 
		\REQUIRE For $k = 0$, set the initial variables $\mathbf{s}_{T,r}^{0},\left\{ \mathbf{h}_{j,r}^{0} \right\}_{j=1}^{{{N}_{T}}L},\left\{ \mathbf{u}_{m,r}^{0} \right\}_{m=1}^{{{N}_{T}}},\left\{ \mathbf{v}_{m,r}^{0} \right\}_{m=1}^{{{N}_{T}}}$,\\
		$\left\{ \mathbf{p}_{j,r}^{0} \right\}_{j=1}^{{{N}_{T}}L},\left\{ \mathbf{q}_{m,r}^{0} \right\}_{m=1}^{{{N}_{T}}},\left\{ \mathbf{d}_{m,r}^{0} \right\}_{m=1}^{{{N}_{T}}}$, and the penalty parameters ${{\rho }_{1}}$, ${{\rho }_{2}}$, and ${{\rho }_{3}}$.\\ 
		\ENSURE The optimal transmit waveforms and receive filter weight pair $\left( {{\mathbf{s}}_{T,opt}},{{\mathbf{w}}_{opt}} \right)$.
		\STATE Let $k = k+1$ and update the $\mathbf{s}_{T}^{k}$ using \eqref{eq:51}.
		\STATE Update the $\left\{ \mathbf{h}_{j,r}^{k} \right\}_{j=1}^{{{N}_{T}}L}$ using \eqref{eq:52a}.
		\STATE Update the $\left\{ \mathbf{u}_{m,r}^{k} \right\}_{m=1}^{{{N}_{T}}}$ using \eqref{eq:52b}.
		\STATE Update the $\left\{ \mathbf{v}_{m,r}^{k} \right\}_{m=1}^{{{N}_{T}}}$ using \eqref{eq:52c}.
		\STATE Compute \\ 
		$\mathbf{p}_{j,r}^{k+1}=\mathbf{p}_{j,r}^{k}+\left[ {{{\mathbf{\bar{E}}}}_{j,r}}\left( \mathbf{s}_{T,r}^{k+1}-{{\mathbf{s}}_{\operatorname{Ref},r}} \right)-\mathbf{h}_{j,r}^{k+1} \right],j=1,2,...,{{N}_{T}}L$.
		\STATE Compute \\
		$\kern 2pt \mathbf{q}_{m,r}^{k+1}=\mathbf{q}_{m,r}^{k}+\left( {{\mathbf{\Sigma }}_{m,r}}\mathbf{s}_{T,r}^{k+1}-\mathbf{u}_{m,r}^{k+1} \right),m=1,2,...,{{N}_{T}}$.
		\STATE Compute \\
		$\kern 2pt \mathbf{d}_{m,r}^{k+1}=\mathbf{d}_{m,r}^{k}+\left( {{{\mathbf{\bar{B}}}}_{m,r}}\mathbf{s}_{T,r}^{k+1}-\mathbf{v}_{m,r}^{k+1} \right),m=1,2,...,{{N}_{T}}$.
		\STATE If the termination conditions are satified, output ${{\mathbf{s}}_\text{T,opt}}=\mathbf{s}_{T}^{k}$ and substitute ${{\mathbf{s}}_\text{T,opt}}$ into \eqref{eq:weig} to obtain ${{\mathbf{w}}_\text{opt}}$. Otherwise, repeat Step 1.
	\end{algorithmic}  
\end{breakablealgorithm}

\subsubsection{Computational complexity and convergence}
\noindent
The computational efficiency of MM-ADMM is evaluated by the number of multiplications required for each iteration.
To optimize $\mathbf{s}_{T}^{k+1}$, the calculation of the matrix inversion has a complexity $\mathcal{O}\left( N_{T}^{3}{{L}^{3}} \right)$.
The update of the remaining auxiliary variables has $\mathcal{O}\left( {{N}_{T}}L \right)$.
In summary, in each iteration of the MM-ADMM, the number of multiplications is $\mathcal{O}\left( {{I}_{2}}N_{T}^{3}{{L}^{3}} \right)$, where $I_2$ denotes number of iterations of the MM-ADMM.
Considering that the MM-ADMM algorithm integrates the MM method and ADMM technology, its convergence can be guaranteed by combining the convergence analysis of the MM method \cite{song2016sequence} and the ADMM method \cite{boyd2011distributed}.

\section{Simulation}\label{sec:d}
\noindent
In this section, numerical examples are presented to verify the effectiveness of the proposed techniques.
The parameters of the simulation scenario and the FDA system are listed in Table \ref{table:taba} and Table \ref{table:tab2}, respectively.
In order to highlight the efficiency of the proposed algorithms, we here consider an extreme scenario where a mainlobe interference (interference $2$) and a normal non-mainlobe interference (interference $1$) coexist and reside in the same radar range resolution unit (similarly, the FDA's range resolution depends on the baseband bandwidth).
In the simulations, the resolution is about $150 \kern 2ptm$.
The orthogonal linear frequency modulation (OLFM) waveform is chosen as the reference waveform, with the $m$th OLFM waveform being \cite{6649991}
\begin{equation}
{{s}_\text{Ref,m}}\left( t \right)={{e}^{j2\pi \left( m-1 \right)\gamma t}}{{e}^{j\pi \left( {{{B}_\text{Ref}}}/{{{T}_{p}}}\; \right){{t}^{2}}}}
\end{equation}
for $m=1,2,...,{{N}_{T}}$, using the bandwidth ${B_{{\text{Ref}}}} = 900 \kern 2pt kHz$ and where $\gamma  = \frac{{{B_\text{Ref}}}}{{{N_T}}}$ represents the frequency interval.
For P-ADMM, we initialize ${\bf{s}}_T^0 = {{\bf{s}}_\text{Ref}}$, where ${{\bf{s}}_\text{Ref}}$ denotes a stack of $N_T$ digital ${s_\text{Ref,m}}\left( t \right)$, ${{\bf{h}}^0} = {\bf{0}}$,
${{\bf{u}}^0} = {\bf{0}}$, and ${\bf{v}}_m^0 = {\bf{0}},m = 1,2,...,{N_T}$.  
For MM-ADMM, we set ${\bf{s}}_T^0 = {{\bf{s}}_\text{Ref}}$, 
${\bf{h}}_j^0 = {\bf{0}},j = 1,2,...,{N_T}L$, ${\bf{u}}_m^0 = {\bf{0}},,m = 1,2,...,{N_T}$, ${\bf{v}}_m^0 = {\bf{0}},m = 1,2,...,{N_T}$, ${\bf{p}}_j^0 = {\bf{0}},j = 1,2,...,{N_T}L$, ${\bf{q}}_m^0 = {\bf{0}},,m = 1,2,...,{N_T}$, and ${\bf{d}}_m^0 = {\bf{0}},m = 1,2,...,{N_T}$.

\begin{table}[t]
	\centering
	\caption{Parameter settings for target and interferences.}
	{\begin{tabular}[l]{@{}lccccc}
			\toprule
			Type &Target  & interference $1$ & interference $2$ \\
			\midrule
			Range &$15.075 \kern 2pt km$  &$14.985 \kern 2pt km$ & $14.970 \kern 2pt km$ \\
			Angle &$20^o$  &$-30^o$ & $20^o$ \\
			Receive spatial frequency &$-0.171$  &$0.250$ & $-0.171$  \\
			Transmit spatial frequency &$0.329$  &$0.15$  & $-0.371$\\
			SNR/INR &$20 \kern 2pt dB$  &$30 \kern 2pt dB$ & $30 \kern 2pt dB$ \\
			\bottomrule
	\end{tabular}}
\label{table:taba}
\end{table}

\begin{table}[t]
	\centering
	\caption{System parameters.}
	{\begin{tabular}[l]{@{}lcllcc}
			\toprule
			Parameter & Value\\
			\midrule
			Number of transmit antennas & $\kern -30pt {N_T}=6$\\
			Number of receive antennas &\kern -30pt ${N_R}=6$\\
			Carrier frequency &$\kern 4pt {f_c}=10 \kern 2pt GHz$ \\
			Reference wavelength &$\kern -8pt \lambda  =  3 \kern 2pt cm$  \\
			Transmit inter-element spacing &\kern -4pt ${d_t}  =  1.5 \kern 2pt cm$  \\
			Receive inter-element spacing &\kern -4.5pt ${d_r}  = 1.5\kern 2pt cm$ \\
			Frequency increment &\kern -4.5pt $\Delta f = 1 \kern 2pt MHz$ \\
			Cutoff frequency of the $m$th LPF &  $\kern -11.5pt {f_{m,lp}} = 900 \kern 2pt kHz$\\
			Radar window &$10 \sim 14\kern 2pt km$ \\
			Pulse duration &${T_p} = 20 \kern 2pt \mu s$\\
			Sampling frequency &${f_s} = 1 \kern 2pt MHz$\\
			Number of samples in radar window & \kern 3pt $\tilde L = \frac{{19 - 10}}{{3{e^8}}} \cdot {f_s} = 30$\\
			Number of samples for each tranmit waveform & \kern 3.1pt $L = {T_p} \cdot {f_s} = 20$\\
			Tolerance of in-band energy &$\kern -7.6pt \begin{array}{l}
				   {\gamma _m} = 0.91\\
				   \kern 4pt m = 1,2,...,{N_T}
				   \end{array}$\\
			\bottomrule
	\end{tabular}}
\label{table:tab2}
\end{table}

\begin{figure}[t]
	\centering
	\includegraphics[width=0.45\textwidth]{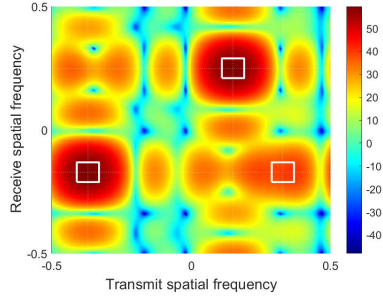}
	\caption{Original power spectral distribution in the transmit-receive spatial frequency domain, where the white rectangles indicate the location of the target $\left( 0,320,-0.171 \right)$, interference 1 $\left( -0.371,-0.171 \right)$, and interference 2 $\left( 0.15,0.250 \right)$.}\label{fig:333}
\end{figure}
\subsection{MainLobe interference suppression}
\noindent
Figure \ref{fig:333} depicts the power spectral distribution of the target and interfering signals in the joint transmit-receive spatial frequency domain. It is computed as 
\begin{equation}
\begin{array}{l}
\kern -10pt P\left( {{f_t},{f_r}} \right) = \\
{{\bf{g}}^H}\left( {{{\bf{s}}_\text{Ref}};{f_t},{f_r}} \right){\bf{\Xi }}\left( {{{\bf{s}}_\text{Ref}};{f_{t,t}},{f_{r,t}},{f_{t,i}},{f_{r,i}}} \right){\bf{g}}\left( {{f_t},{f_r},{{\bf{s}}_\text{Ref}}} \right)
\end{array}
\end{equation}
where 
\begin{subequations}
	\begin{equation}
	\kern -110pt \mathbf{g}\left( {{\mathbf{s}}_\text{Ref}};{{f}_{t}},{{f}_{r}} \right)=\mathbf{\Lambda }\left( {{f}_{t}},{{f}_{r}} \right){{\mathbf{s}}_\text{Ref}}
	\end{equation}
	\begin{equation}
	\begin{array}{l}
	{\bf{\Xi }}\left( {{{\bf{s}}_\text{Ref}};{f_{t,t}},{f_{r,t}},{f_{t,i}},{f_{r,i}}} \right)\\
	\kern 63pt = {\mathop{\rm SNR}\nolimits}  \cdot {\bf{g}}\left( {{{\bf{s}}_\text{Ref}};{f_{t,t}},{f_{r,t}}} \right){{\bf{g}}^H}\left( {{{\bf{s}}_\text{Ref}};{f_{t,t}},{f_t}} \right)\\
	\kern 40pt + \sum\limits_{i = 1}^{\Upsilon} {\left\{ {{\rm{IN}}{{\rm{R}}_i} \cdot {\bf{g}}\left( {{{\bf{s}}_\text{Ref}};{f_{t,i}},{f_{r,t}}} \right){{\bf{g}}^H}\left( {{{\bf{s}}_\text{Ref}};{f_{t,i}},{f_t}} \right)} \right\}} 
	\end{array}
	\end{equation}
\end{subequations}
with $\mathbf{\Lambda }\left( {{f}_{t}},{{f}_{r}} \right)$ being given in \eqref{eq:38c}.
In a traditional MIMO radar system, the targets and interferences will be diagonally distributed in the joint transmit-receive domain.
It can be seen that, for FDA, the interferences and the target are separated due to the different transmit spatial frequencies.

\begin{figure}[t]
	\centering
	\subfigure[]{
		\begin{minipage}[b]{0.45\textwidth}
			\includegraphics[width=1\textwidth]{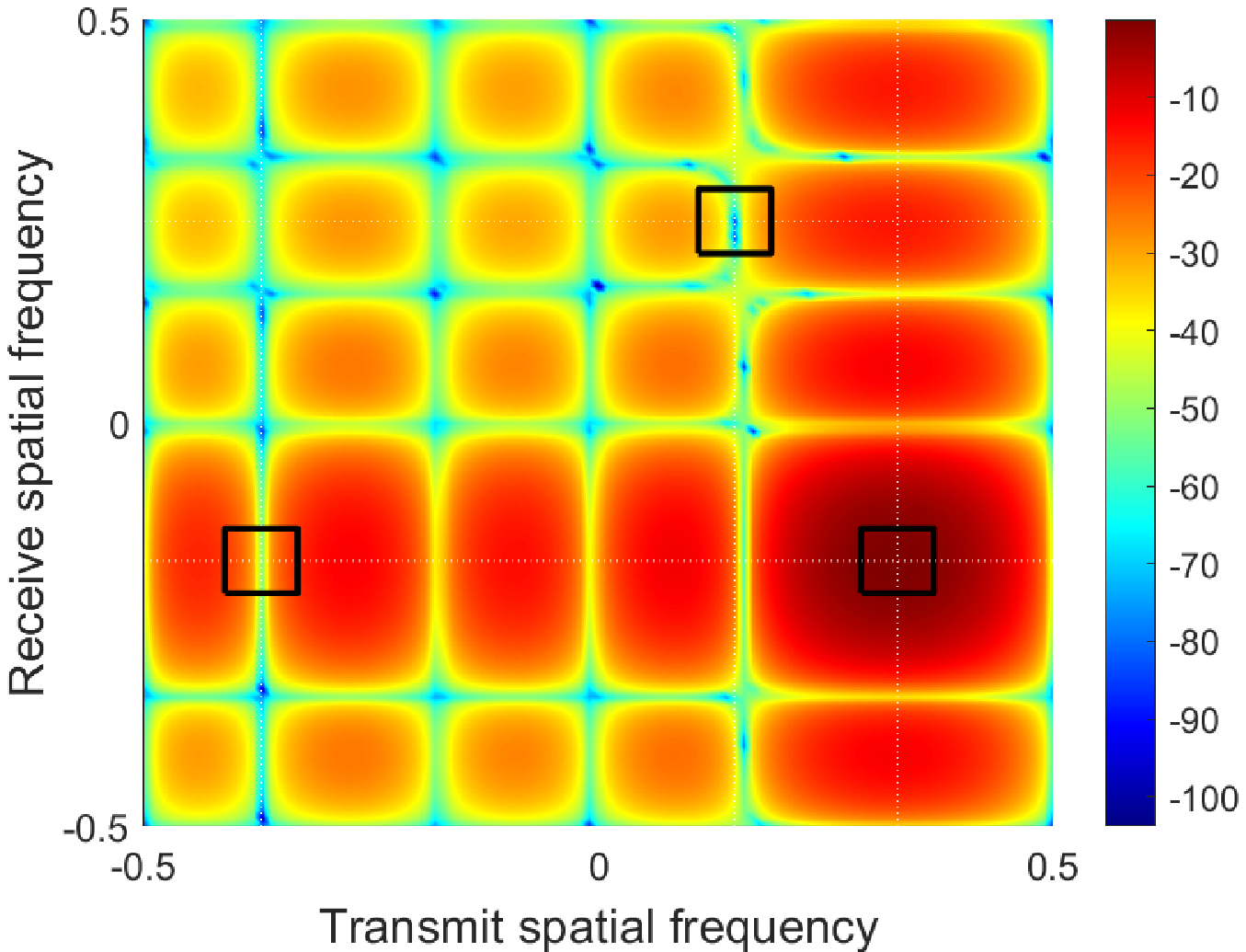}\label{fig:444a}
		\end{minipage}
	}
	\subfigure[]{
		\begin{minipage}[b]{0.45\textwidth}
			\includegraphics[width=1\textwidth]{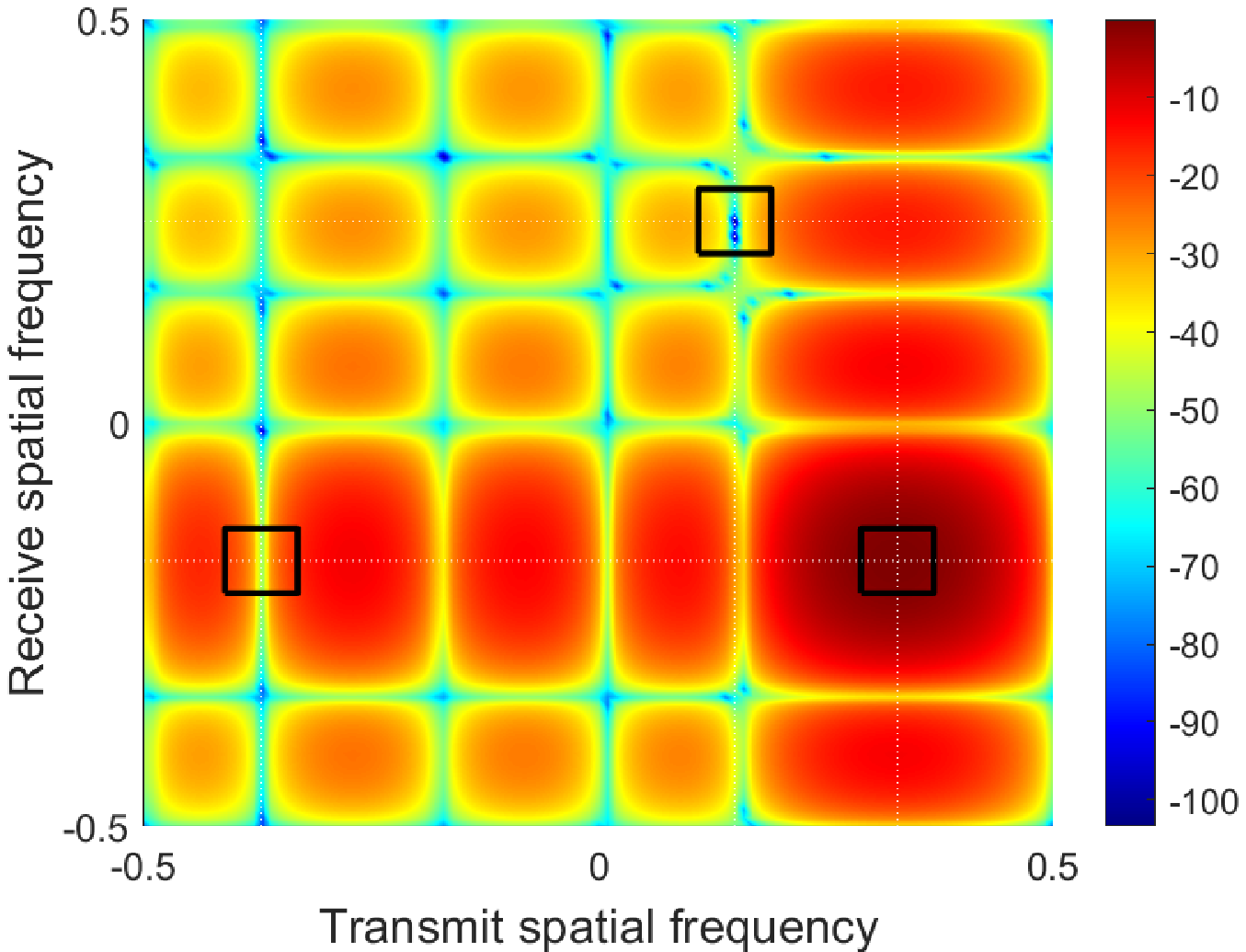}\label{fig:444b}
		\end{minipage}
	}
	\caption{Output power spectrum distribution where the black rectangles indicate the location of the target and interferences. (a) $\varepsilon =0.2$. (b) $\varepsilon =1$.} \label{fig:444}
\end{figure}

Considering different similarity levels $\varepsilon $, Figure \ref{fig:444} illustrates the spectral distribution of the interferences and the target in the joint transmit and receive domain after designing the transmit waveform and receive filter weights. The output distribution is computed from
\begin{equation}
{P_\text{output}}\left( {{f_t},{f_r}} \right) = {\left| {{\bf{w}}_\text{opt}^H\mathbf{\Lambda }\left( {{f}_{t}},{{f}_{r}} \right){{\bf{s}}_\text{T,opt}}} \right|^2}
\end{equation}
where ${{\bf{w}}_\text{opt}}$ and ${{{\bf{s}}_\text{T,opt}}}$ are obtained by the MM-ADMM algorithm.
The results demonstrate that the proposed algorithm is able to focus on the target with the correct transmit and receive spatial frequency pair and form notches at both the mainlobe and normal interference locations.

\begin{figure}[t]
	\centering
	\subfigure[]{
		\begin{minipage}[b]{0.45\textwidth}
			\includegraphics[width=1\textwidth]{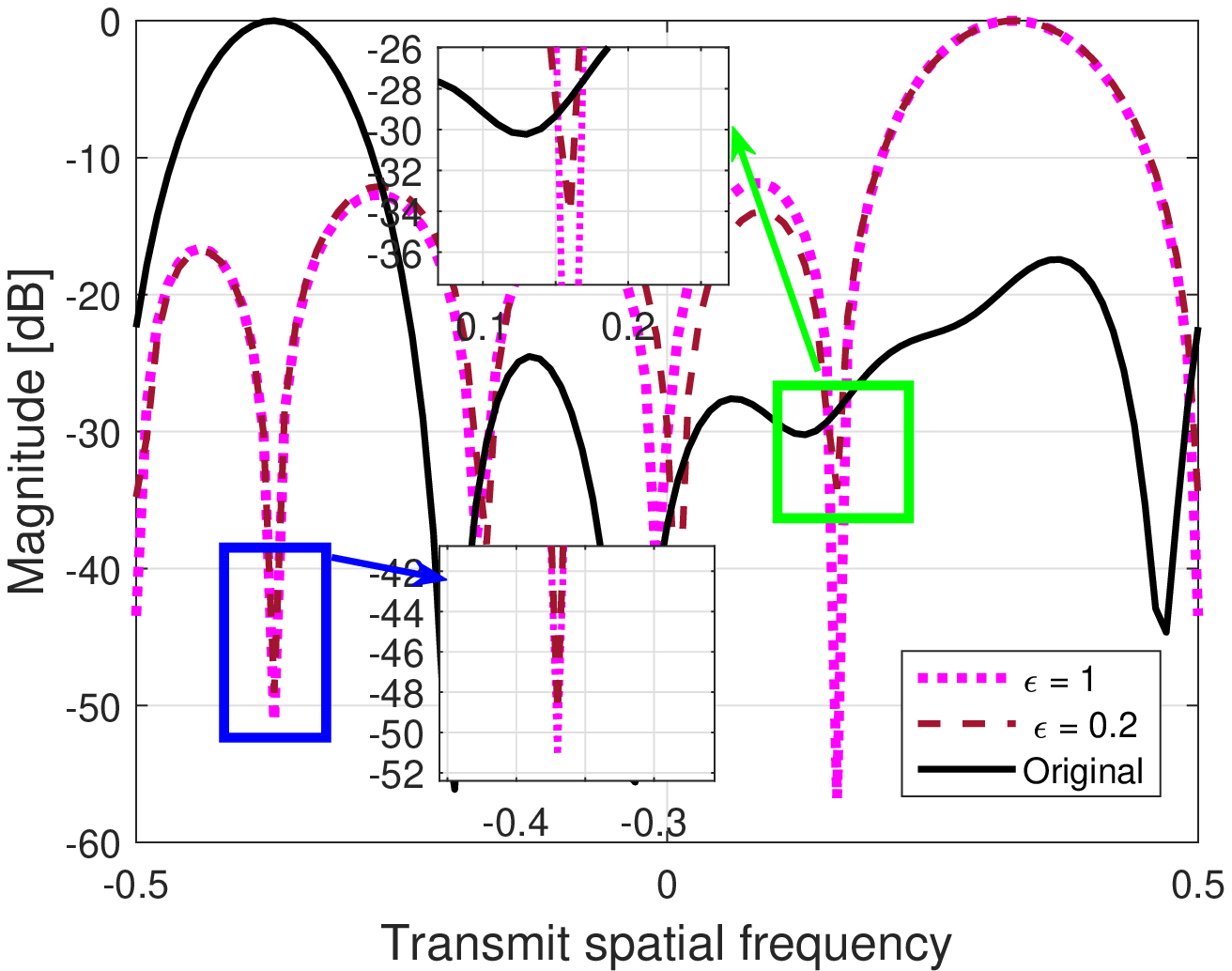}
		\end{minipage}
	}
	\subfigure[]{
		\begin{minipage}[b]{0.45\textwidth}
			\includegraphics[width=1\textwidth]{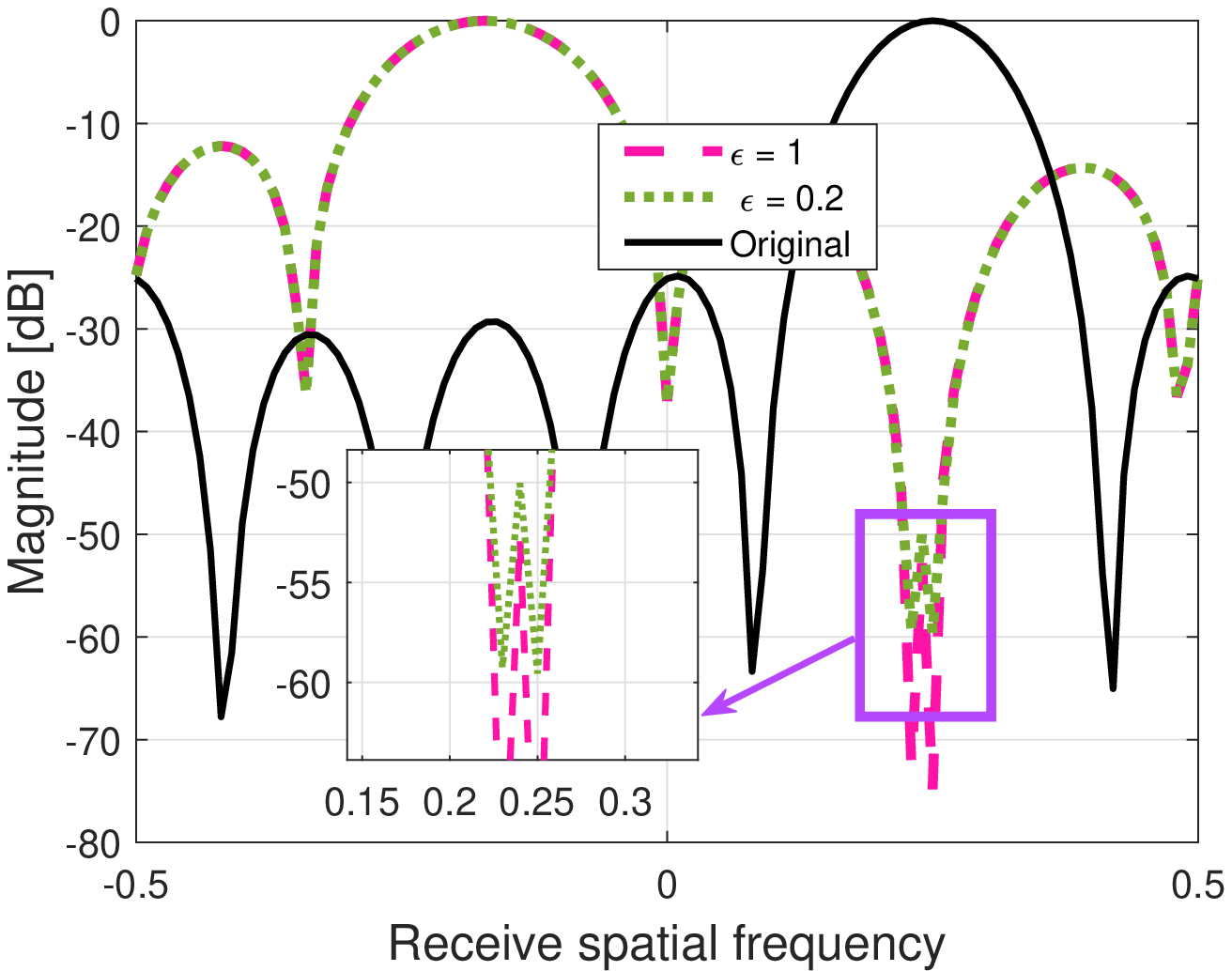}
		\end{minipage}
	}
	\caption{Cut of the output spectrum at the (a) receive spatial frequency  ${f_{r,2}} =  -0.171$ and (b) transmit spatial frequency  ${f_{t,2}} = 0.15$.}\label{fig:555} 	
\end{figure}

\begin{figure}[t]
	\centering
	\includegraphics[width=0.45\textwidth]{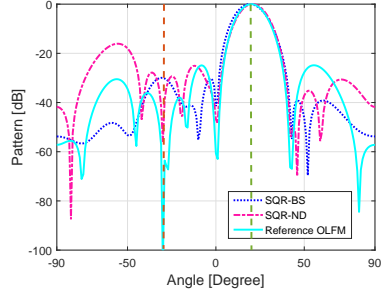}
	\caption{Output beampattern of MIMO radar where the orange and green dashed lines indicate the azimuths of interference $1$ and interference $2$, respectively.}\label{fig:666}
\end{figure}

In order to illustrate the effect of the similarity level $\varepsilon $ on the suppression performance, Figure \ref{fig:555} shows the cut of the output spectrum at the receive spatial frequency ${f_{r,1}} = -0.171$ and the transmit spatial frequency ${f_{t,2}} =  0.15$.
As expected, the larger the similarity level $\varepsilon $, the wider the selection of waveforms, and the deeper the resulting notches at the the location of the interference signals.
In comparison, the output beampattern of the MIMO radar is shown in Figure \ref{fig:666}. Here, the transmit waveform and the optimal filter weights of the MIMO radar are obtained by the SQR-BS and SQR-ND algorithms \cite{7450660}.
It is clear from Figure \ref{fig:666} that deep notches are generated as a result of the MIMO radar processing at the location of the normal interference (interference $1$).
However, the figure does not clearly indicate whether the MIMO weight have been able to suppress interference $2$, being at the same azimuth angle as the target.
One conjecture is that the MIMO radar suppresses interference $2$ while focusing the echo energy at the target location.
In order to determine if this is so, we compute the resulting output SINR, thereby obtaining a performance comparison of the FDA and the MIMO radars in terms of output SINR.

\begin{figure}[t]
	\centering
	\includegraphics[width=0.45\textwidth]{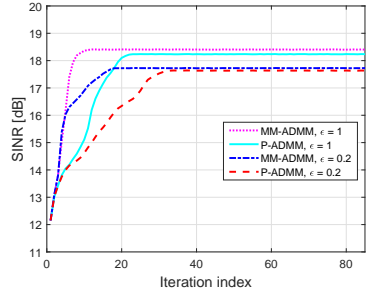}
	\caption{FDA output SINR values versus the iteration number.}\label{fig:777}
\end{figure}
\subsection{Output SINR}
\noindent
Figure \ref{fig:777} shows the output SINR behaviors versus the iteration number for different similarity levels $\varepsilon $.
The intercept (initial iteration) of the SINR curve on the ordinate axis corresponds to the FDA performance using a conventional receiver.
From the figure, it can be noted that the output SINR after the MVDR beamforming is about $12.15 \kern 2pt dB$.
It is also clear that the P-ADMM and MM-ADMM designs result in a transmit waveform which yields a gradually improving performance as the algorithms converge.
According to parameters in Table \ref{table:taba}, the output SINR should reach the upper bound of $20 \kern 2pt dB$ without interferences.
Therefore, it can be seen that the two proposed algorithms are able to improve the SINR by about $6 \kern 2pt dB$, making the resulting performance gap less than $2 \kern 2pt dB$.
Since MM-ADMM has an analytical solution at each iteration, its complexity is significantly lower than that of P-ADMM.
Comparing the results of the MIMO radar for normal signal-dependent interference \cite{6649991,7450660}, it can be seen that the output SINR achievable by using with waveform design is slightly smaller.
This may be due to the made approximations, or it is because the FDA achieves interference suppression by acting on both the transmit and the receive spatial frequencies.
Finally, it can be noted that the two algorithms perform similarly, with the achievable SINR being related to the used similarity level, $\varepsilon $. The larger the similarity level, the larger the output SINR, as the use of a larger similarity level implies the use of a wider selection of waveforms.

\begin{figure}[t]
	\centering
	\includegraphics[width=0.45\textwidth]{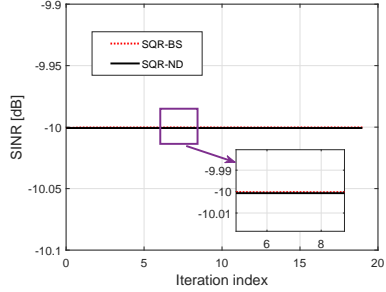}
	\caption{MIMO output SINR values versus the iteration number.}\label{fig:888}
\end{figure}

In comparison, the results from the MIMO radar designs are shown in Figure \ref{fig:888}, indicating that the output SINR of the MIMO radar design stabilizes at $-10 \kern 2pt dB$.
As shown in Table \ref{table:taba}, the INR of both interferers are $0 \kern 2pt dB$, whereas the SNR of the target is $20 \kern 2pt dB$. Thus, comparing the results in Figures \ref{fig:666} and \ref{fig:888}, one may conclude that the MIMO radar suppresses interference $1$ but not the mainlobe interference $2$.
This as when only interference $2$ and the target are present, the SINR $-10 \kern 2pt dB$, which corresponds to the simulation result shown in Figure \ref{fig:888}.
We can therefore conclude that the MIMO radar is unable to suppress the influence of the mainlobe interference. The output SINR of the PA array is similar to that of the MIMO radar.

\subsection{Pulse compression}
\begin{figure}[htbp]
	\centering
	\includegraphics[width=0.45\textwidth]{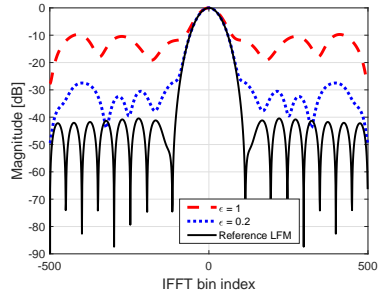}
	\caption{Pulse compression profile of the designed waveform.}\label{fig:999}
\end{figure}

\noindent
Finally, the pulse compression property of the designed waveforms is considered, with the resulting simulation results being shown in Figure \ref{fig:999}.
For comparison, the property of the reference OLFM waveform is also included in the simulation.

Here, we have use the excitation waveform designed by the proposed MM-ADMM algorithm for transmit antennas $1$. For other waveforms, the simulation results are formed similarly, with the simulation results being obtained using a frequency-domain matched filter, with an $L$-point Hamming window being used to reduce the side lobes and a $1000$-point inverse fast fourier fransform (IFFT) operation to increase the sampling rate.
The results show that as the similarity level $\varepsilon $ gradually decreases, the width of the mainlobe becomes narrower and the sidelobe becomes lower.
Combined with the results shown in Figure \ref{fig:777}, showing that a larger similarity level corresponds to a larger achievable output SINR, it may be concluded that, the choice of similarity value requires a trade-off between pulse compression and the SINR performance.

\section{Conclusion}\label{sec:e}
In this paper, we have designed an FDA transmit waveform improving the output SINR of the system, allowing for the significant advantages of FDA in enabling mainlobe interference cancellation.
Since the current FDA receiver based on multi-carrier matched filtering cannot effectively explore the waveform diversity of FDA, we present an FDA receiver based on multi-channel low pass filtering to better separate the transmit waveform at the receiving side.
Unlike a MIMO radar, the FDA received signal is range-angle dependent, implying that existing FDA signal modeling ignoring the propagation delay of different interferences is unable to fully exploit the FDA structure.
To enable this, we formulate a complete FDA receiver output signal model considering the designed multi-channel low pass filtering receiver and propagation delay.
Next, we employ an MVDR beamforming formulation to form the optimal receive filter weights.
The resulting FDA transmit waveform design problem of the output SINR maximization can be formulated as an optimization problem involving a non-convex objective function with multiple non-convex constraints.
We introduce two iterative approximative optimization algorithms, termed the P-ADMM and MM-ADMM, to solve the resulting problem.
Finally, we evaluate the performance of the designed waveforms obtained by the proposed algorithms in terms of mainlobe interference supression, output SINR, and pulse compression, demonstrating the advantages of the FDA radar in comparison with corresponding MIMO radar.
The results show that, similar to the case for MIMO radar, the output SINR of an FDA can be significantly improved using waveform design.
In conclusion, we note that the proposed designed receiver and the established output signal model are only applicable to the case where the bandwidth of the transmit waveform is smaller than the frequency increment, suggesting an interesting area for further development.
\newline

\appendix
\section{}
\label{sec:appen}
Define the proximal operator ${{\operatorname{prox}}_{\rho f}}:{{\mathbb{R}}^{n}}\to {{\mathbb{R}}^{n}}$ as
\begin{equation}
{{\mathop{\rm prox}\nolimits} _\rho }\left( {{\bf{v}};D\left( {\bf{x}} \right)} \right) = \mathop {\arg \min }\limits_{{\bf{x}} \in D\left( {\bf{x}} \right)} {\mkern 1mu} \frac{\rho }{2}{\left\| {{\bf{x}} - {\bf{v}}} \right\|^2}.
\end{equation}
Then,
\begin{equation}
{{\mathbf{x}}_{opt}}={{\operatorname{prox}}_{\rho }}\left( \mathbf{v} \right)={{\prod }_{D\left( \mathbf{x} \right)}}\left( \mathbf{v} \right)
\end{equation}
where ${\prod _{D\left( {\bf{x}} \right)}}\left( {\bf{v}} \right)$ is the Euclidean projection onto the set ${D\left( {\bf{x}} \right)}$.
Taking problem \ref{49b} as an example, it may then be equivalently expressed as
\begin{equation}
\begin{array}{l}
{\bf{h}}_{j,r}^{k + 1} = \\
{{\mathop{\rm prox}\nolimits} _\rho }\left( {{h_j}\left( {{\bf{s}},{\bf{p}}} \right);D\left( {{{\bf{h}}_{j,r}}} \right)} \right) = {{\bf{\Pi }}_{D\left( {{{\bf{h}}_{j,r}}} \right)}}\left( {{h_j}\left( {{\bf{s}}_{T,r}^{k + 1},{\bf{p}}_{j,r}^k} \right)} \right).
\end{array}
\end{equation}
where ${{h}_{j}}\left( \mathbf{s},\mathbf{p} \right)={{{\mathbf{\bar{E}}}}_{j,r}}\left( \mathbf{s}_{T,r}^{k+1}-{{\mathbf{s}}_{\operatorname{Ref},r}} \right)+\mathbf{p}_{j,r}^{k}$ and $D\left( {{{\bf{h}}_{j,r}}} \right) = \left\{ {{{\bf{h}}_{j,r}}\left| {\left\| {{{\bf{h}}_{j,r}}} \right\|_2^2 \le \varepsilon^2 ,\varepsilon  \ge 0} \right.} \right\}$.
Thus,
\begin{equation}
{\bf{h}}_{j,r}^{k + 1} = \left\{ {\begin{array}{*{20}{c}}
	{{h_j}\left( {{\bf{s}},{\bf{p}}} \right)}&{\left\| {{h_j}\left( {{\bf{s}},{\bf{p}}} \right)} \right\|_2^2 \le \varepsilon^2 }\\
	{\frac{{{h_j}\left( {{\bf{s}},{\bf{p}}} \right) \varepsilon  }}{{\left\| {{h_j}\left( {{\bf{s}},{\bf{p}}} \right)} \right\|_2^2}}}&{\left\| {{h_j}\left( {{\bf{s}},{\bf{p}}} \right)} \right\|_2^2 \ge \varepsilon^2 }
	\end{array}} \right.
\end{equation}
where ${{h}_{j}}\left( \mathbf{s},\mathbf{p} \right)={{{\mathbf{\bar{E}}}}_{j,r}}\left( \mathbf{s}_{T,r}^{k+1}-{{\mathbf{s}}_{\operatorname{Ref},r}} \right)+\mathbf{p}_{j,r}^{k}$.
The update of other auxiliary variables are may be formed similarly to the update of ${{\mathbf{h}}_{j,r}}$.
This completes the proof.

\section{}
\label{sec:appenRx}
Combining the output signals of $N_R$ receive antennas into a ${{\text{N}}_{T}}\times {{\text{N}}_{R}}$-dimensional space-time matrix yields
\begin{equation}
\begin{array}{*{35}{l}}
\mathbf{R}\left( t \right)=\left[ {{\mathbf{r}}_{1}}\left( t \right),{{\mathbf{r}}_{2}}\left( t \right),...,{{\mathbf{r}}_{{{N}_{R}}}}\left( t \right) \right]  \\
\kern 24pt =\varsigma \left( {{r}_{t}},{{\theta }_{t}} \right)\text{diag}\left\{ {{\mathbf{a}}_{\text{T}}}\left( {{r}_{t}},{{\theta }_{t}} \right) \right\}\mathbf{s}\left( t-\frac{2{{r}_{t}}}{c} \right)\mathbf{a}_{\text{R}}^{T}\left( {{\theta }_{t}} \right)  \\
\end{array}
\end{equation}
where 
\begin{equation}
{{\mathbf{a}}_{\text{R}}}\left( {{\theta }_{t}} \right)= {\left[ {1,{e^{ - j2\pi \frac{{{d_r}\sin \theta_t }}{\lambda }}},...,{e^{ - j2\pi \frac{{\left( {{N_R} - 1} \right){d_r}\sin \theta_t }}{\lambda }}}} \right]^T}
\end{equation}
denotes the FDA receive steering vector.
It is worth noting that the transmit waveform vector ${\bf{s}}\left( t \right)$ has thus been separated, and that the received signal matrix ${\bf{R}}\left( t \right)$ depends on both the range and angle to the target.

As a consequence, including the presence of signal-dependent interference and Gaussian noise, the ${{\text{N}}_{T}}\times {{\text{N}}_{R}}$-dimensional FDA receive signal matrix may be expressed as
\begin{equation}
\begin{aligned}
& {{\mathbf{R}}_\text{FDA}}\left( t \right)=\varsigma \left( {{r}_{t}},{{\theta }_{t}} \right)\text{diag}\left\{ {{\mathbf{a}}_{\text{T}}}\left( {{r}_{t}},{{\theta }_{t}} \right) \right\}\mathbf{s}\left( t-\frac{2{{r}_{t}}}{c} \right)\mathbf{a}_{\text{R}}^{T}\left( {{\theta }_{t}} \right) \\ 
& \kern -3pt +\sum\limits_{i=1}^{\Upsilon}{\varsigma \left( {{r}_{i}},{{\theta }_{i}} \right)\text{diag}\left\{ {{\mathbf{a}}_{\text{T}}}\left( {{r}_{i}},{{\theta }_{i}} \right) \right\}\mathbf{s}\left( t-\frac{2{{r}_{i}}}{c} \right)\mathbf{a}_{\text{R}}^{T}\left( {{\theta }_{i}} \right)}+\mathbf{N}\left( t \right) \\ 
\end{aligned}
\end{equation}
where $\Upsilon$ represents the number of interferences, ${\bf{N}}\left( t \right)$ is the noise matrix,
and $\varsigma \left( {{r_i},{\theta _i}} \right)$ denotes the complex amplitude of the $i$th interference source.
We here term an interference which has the same azimuth as the target a mainlobe interference.
The ${{\text{N}}_{T}} {{\text{N}}_{R}}$-dimensional FDA space-time snapshot can be expressed as
\begin{equation}
\begin{aligned}
& {{{\mathbf{\bar{r}}}}_\text{FDA}}\left( t \right)  
=\varsigma \left( {{r}_{t}},{{\theta }_{t}} \right)\left[ {{\mathbf{a}}_{\text{R}}}\left( {{\theta }_{t}} \right)\otimes \text{diag}\left\{ {{\mathbf{a}}_{\text{T}}}\left( {{r}_{t}},{{\theta }_{t}} \right) \right\} \right]\mathbf{s}\left( t-\frac{2{{r}_{t}}}{c} \right) \\ 
& \kern 40pt +\sum\limits_{i=1}^{\Upsilon}{\left\{ \begin{array}{*{35}{l}}
	\left[ {{\mathbf{a}}_{\text{R}}}\left( {{\theta }_{i}} \right)\otimes \text{diag}\left\{ {{\mathbf{a}}_{\text{T}}}\left( {{r}_{i}},{{\theta }_{i}} \right) \right\} \right]  \\
	\cdot \varsigma \left( {{r}_{i}},{{\theta }_{i}} \right)\cdot \mathbf{s}\left( t-\frac{2{{r}_{i}}}{c} \right)  \\
	\end{array} \right\}}+\mathbf{\bar{n}}\left( t \right) \\ 
\end{aligned}
\end{equation}
where ${\bf{\bar n}}\left( t \right) = {\mathop{\rm vec}\nolimits} \left\{ {{\bf{N}}\left( t \right)} \right\}$ is assumed to be in both spatially and temporally white.

Since the FDA has a unique range-dependent steering vector, the radar range gate has to be taken into account when modeling the FDA signal,
as illustrated in Figure \ref{fig:222}, considering the fact that interference echos from different range gates require different propagation times to reach the radar receiver. In Figure \ref{fig:222}, $\tilde L$ denotes the length of the radar window, whereas the number of samples for each transmit waveform pulse is $L$. Denote the range gates of the target
and of the $i$th interference $l^t$ and $l^i$, respectively.
Then, the FDA snapshot ${{{\bf{\bar r}}}_\text{FDA}}\left( t \right)$ after analog-to-digital conversion (ADC) can be written as
\begin{equation}
\begin{array}{l}
\kern -3pt {{{\bf{\bar r}}}_\text{FDA}}\left( l \right)  =\varsigma \left( {{r_t},{\theta _t}} \right)\left[ {{{\bf{a}}_{\mathop{\rm R}\nolimits} }\left( {{\theta _t}} \right) \otimes {\mathop{\rm diag}\nolimits} \left\{ {{{\bf{a}}_{\mathop{\rm T}\nolimits} }\left( {{r_t},{\theta _t}} \right)} \right\}} \right]{\bf{c}}\left( {{l^t,l}} \right) \\
\kern -5pt + \sum\limits_{i = 1}^\Upsilon {\left\{ {\varsigma \left( {{r_i},{\theta _i}} \right)\left[ {{{\bf{a}}_{\mathop{\rm R}\nolimits} }\left( {{\theta _i}} \right) \otimes {\mathop{\rm diag}\nolimits} \left\{ {{{\bf{a}}_{\mathop{\rm T}\nolimits} }\left( {{r_i},{\theta _i}} \right)} \right\}} \right]{\bf{c}}\left( {{l^i,l}} \right)} \right\}} + {\bf{\bar n}}\left( l \right) \\
\end{array}
\end{equation}
for $l = 1,2,...,\tilde L$, where ${\bf{c}}\left( {{l^t,l}} \right)$ and ${\bf{c}}\left( {{l^i,l}} \right)$ represent the digital signals associated with $\mathbf{s}\left( t-\frac{2{{r}_{t}}}{c} \right)$ and $\mathbf{s}\left( t-\frac{2{{r}_{i}}}{c} \right)$, respectively.
Concatenating the responses of all samples into a ${{\text{N}}_{T}}{{\text{N}}_{R}}\times \tilde{L}$-dimensional matrix yields
\begin{equation}
\begin{aligned}
& \kern -4pt {{{\mathbf{\bar{R}}}}_\text{FDA}}=\left[ \begin{matrix}
{{{\mathbf{\bar{r}}}}_\text{FDA}}\left( 1 \right) & {{{\mathbf{\bar{r}}}}_\text{FDA}}\left( 2 \right) & ... & {{{\mathbf{\bar{r}}}}_\text{FDA}}\left( {\tilde{L}} \right)  \\
\end{matrix} \right] \\ 
& \kern 19pt =\varsigma \left( {{r}_{t}},{{\theta }_{t}} \right)\left[ {{\mathbf{a}}_{\text{R}}}\left( {{\theta }_{t}} \right)\otimes \text{diag}\left\{ {{\mathbf{a}}_{\text{T}}}\left( {{r}_{t}},{{\theta }_{t}} \right) \right\} \right]\mathbf{SK}\left( {{l}^{t}} \right) \\ 
& +\sum\limits_{i=1}^{\Upsilon}{\left\{ \varsigma \left( {{r}_{i}},{{\theta }_{i}} \right)\left[ {{\mathbf{a}}_{\text{R}}}\left( {{\theta }_{i}} \right)\otimes \text{diag}\left\{ {{\mathbf{a}}_{\text{T}}}\left( {{r}_{i}},{{\theta }_{i}} \right) \right\} \right]\mathbf{SK}\left( {{l}^{i}} \right) \right\}} +\mathbf{\bar{N}} \\ 
\end{aligned}
\end{equation}
where $\mathbf{\bar{N}}=\left[ \mathbf{n}\left( 1 \right),\mathbf{n}\left( 2 \right),...,\mathbf{n}\left( {\tilde{L}} \right) \right]\in {{\mathbb{C}}^{{{N}_{T}}{{N}_{R}}\times \tilde{L}}}$
represents the noise matrix,
${\bf{SK}}\left( {{l^j}} \right) \in {{\mathbb{C}}^{{N_T} \times \tilde L}} = \left[ {{\bf{c}}\left( {l^j},1 \right),{\bf{c}}\left( {l^j},2 \right),...,{\bf{c}}\left( {l^j},{\tilde L} \right)} \right]$
denotes the "shift" transmit waveform matrix with
${\bf{S}} \in {{\mathbb{C}}^{{N_T} \times L}} = \left[ {{\bf{s}}\left( 1 \right),{\bf{s}}\left( 2 \right),...,{\bf{s}}\left( L \right)} \right]$
being the transmit waveform matrix and
$\mathbf{K}\left( {{l}^{j}} \right)\in {{\mathbb{C}}^{L\times \tilde{L}}}$
a "shift" matrix.
As illustrated in Figure \ref{fig:222}, for interference $1$ arriving earlier than the start of the radar sampling, this yields \cite{8356676}
\begin{equation}
\mathbf{K}\left( {{l}^{j}} \right)=\left[ \begin{matrix}
{{\mathbf{0}}_{{{l}^{j}}\times \left( \tilde{L}-{{l}^{j}} \right)}} & {{\mathbf{I}}_{{{l}^{j}}}}  \\
{{\mathbf{0}}_{\left( L-{{l}^{j}} \right)\times \left( \tilde{L}-{{l}^{j}} \right)}} & {{\mathbf{0}}_{\left( L-{{l}^{j}} \right)\times {{l}^{j}}}}  \\
\end{matrix} \right].
\end{equation}
Similarly, for the target and interference $2$,
\begin{equation}
{\mathbf{K}}\left( {{l^j}} \right) = \left[ {\begin{array}{*{20}{c}}
	{{{\mathbf{0}}_{L \times {l^j}}}}&{{{\mathbf{I}}_L}}&{{{\mathbf{0}}_{L \times \left( {\tilde L - {l^j} - L} \right)}}} 
	\end{array}} \right],
\end{equation}
whereas for interference $3$ arriving later than the end of radar sampling,
\begin{equation}
\mathbf{K}\left( {{l}^{j}} \right)=\left[ \begin{matrix}
{{\mathbf{0}}_{\left( L-{{l}^{j}} \right)\times \left( \tilde{L}-{{l}^{j}} \right)}} & {{\mathbf{0}}_{\left( L-{{l}^{j}} \right)\times {{l}^{j}}}}  \\
{{\mathbf{0}}_{{{l}^{j}}\times \left( \tilde{L}-{{l}^{j}} \right)}} & {{\mathbf{I}}_{{{l}^{j}}}}  \\
\end{matrix} \right].
\end{equation}

\section{}
\label{sec:appenMMx}
The real-valued version of problem $P_4$ can be expressed as
\begin{equation}
{{{P}}_{5}}\left\{ \begin{matrix}
\underset{{{\mathbf{s}}_{T,r}}}{\mathop{\min }}\, & 2{{\mathbf{s}}_{T,r}}{{\mathbf{P}}_{r}}\left( {{{\mathbf{\tilde{S}}}}_{T}} \right)\mathbf{s}_{T,r}^{H}-2\mathbf{z}_{r}^{T}\left( {{{\mathbf{\tilde{s}}}}_{T}};{{{\mathbf{\tilde{S}}}}_{T}} \right){{\mathbf{s}}_{T,r}}  \\
\operatorname{s}.t. & \left\{ \begin{matrix}
{{\left( {{\mathbf{s}}_{T,r}}-{{\mathbf{s}}_{\operatorname{Ref},r}} \right)}^{T}}{{{\mathbf{\bar{E}}}}_{j,r}}\left( {{\mathbf{s}}_{T,r}}-{{\mathbf{s}}_{\operatorname{Ref},r}} \right)\le \varepsilon^2  \\
\begin{aligned}
& \mathbf{s}_{T,r}^{H}{{\mathbf{\Sigma }}_{m,r}}{{\mathbf{s}}_{T,r}}=\frac{1}{{{N}_{T}}} \\ 
& \mathbf{s}_{T,r}^{H}{{\mathbf{B}}_{m,r}}{{\mathbf{s}}_{T,r}}\ge \frac{{{\gamma }_{m}}}{{{N}_{T}}} \\ 
\end{aligned}  \\
\end{matrix} \right.  \\
\end{matrix} \right.,
\end{equation}
where ${{\mathbf{s}}_{T,r}}$, ${{\mathbf{P}}_{r}}\left( {{{\mathbf{\tilde{S}}}}_{T}} \right)$, ${{\mathbf{z}}_{r}}\left( {{{\mathbf{\tilde{s}}}}_{T}};{{{\mathbf{\tilde{S}}}}_{T}} \right)$, ${{\mathbf{s}}_{\operatorname{Ref},r}}$, ${{{\mathbf{\bar{E}}}}_{j,r}}$, ${{\mathbf{\Sigma }}_{m,r}}$, and ${{\mathbf{B}}_{m,r}}$ are the real-valued forms of ${{\mathbf{s}}_{T}}$, $\mathbf{P}\left( {{{\mathbf{\tilde{S}}}}_{T}} \right)$, $\mathbf{z}\left( {{{\mathbf{\tilde{s}}}}_{T}};{{{\mathbf{\tilde{S}}}}_{T}} \right)$, ${{\mathbf{s}}_{\operatorname{Ref}}}$, ${{{\mathbf{\bar{E}}}}_{j}}$, ${{\mathbf{\Sigma }}_{m}}$, and ${{\mathbf{B}}_{m}}$, respectively.
Next, we introduce the auxiliary variables $\left\{ {{\mathbf{h}}_{j,r}} \right\}_{j=1}^{{{N}_{T}}L}$, $\left\{ {{\mathbf{u}}_{m,r}} \right\}_{m=1}^{{{N}_{T}}}$, and $\left\{ {{\mathbf{v}}_{m,r}} \right\}_{m=1}^{{{N}_{T}}}$ to transform $P_5$ into
\begin{equation}
{{{P}}_{6}}\left\{ \begin{matrix}
\underset{{{\mathbf{s}}_{T,r}}}{\mathop{\min }}\, & 2{{\mathbf{s}}_{T,r}}{{\mathbf{P}}_{r}}\left( {{{\mathbf{\tilde{S}}}}_{T}} \right)\mathbf{s}_{T,r}^{H}-2\mathbf{z}_{r}^{T}\left( {{{\mathbf{\tilde{s}}}}_{T}};{{{\mathbf{\tilde{S}}}}_{T}} \right){{\mathbf{s}}_{T,r}}  \\
\operatorname{s}.t. & \left\{ \begin{matrix}
\begin{aligned}
& \kern 2pt {{{\mathbf{\bar{E}}}}_{j,r}}\left( {{\mathbf{s}}_{T,r}}-{{\mathbf{s}}_{\operatorname{Ref},r}} \right)={{\mathbf{h}}_{j,r}} \\ 
& \kern 25pt \left\| {{\mathbf{h}}_{j,r}} \right\|_{2}^{2}\le \varepsilon^2\\ 
& \kern 25pt {{\mathbf{\Sigma }}_{m,r}}{{\mathbf{s}}_{T,r}}={{\mathbf{u}}_{m,r}} \\ 
\end{aligned}  \\
\begin{aligned}
& \kern 18pt \left\| {{\mathbf{u}}_{m,r}} \right\|_{2}^{2}=\frac{1}{{{N}_{T}}} \\ 
& \kern 16pt {{{\mathbf{\bar{B}}}}_{m,r}}{{\mathbf{s}}_{T,r}}={{\mathbf{v}}_{m,r}} \\ 
& \kern 20pt \left\| {{\mathbf{v}}_{m,r}} \right\|_{2}^{2}\ge \frac{{{\gamma }_{m}}}{{{N}_{T}}} \\ 
\end{aligned}  \\
\end{matrix} \right.  \\
\end{matrix} \right.,
\end{equation}
where ${{{\mathbf{\bar{B}}}}_{m,r}}$ is the real-valued form of ${{{\mathbf{\bar{B}}}}_{m}}$ and represents a block diagonal matrix, with its $m$th diagonal block being an identity matrix ${{{\mathbf{\bar{H}}}}_{m}}$.
with ${\bf{\bar { H}}}_m$ being the Cholesky decomposition of ${{\bf{{H}}}_m}$ and
${{\bf{{H}}}_m} = {\bf{\bar { H}}}_m^H{{{\bf{\bar {H}}}}_m}$.
The resulting scaled augmented Lagrangian function of $P_6$ is given by
\begin{equation}
\begin{aligned}
& {{\overline{\mathsf{\mathbb{L}}}}_{\text{MM-ADMM}}}\left( \begin{aligned}
& {{\mathbf{s}}_{T,r}},\left\{ {{\mathbf{h}}_{j,r}} \right\}_{j=1}^{{{N}_{T}}L},\left\{ {{\mathbf{u}}_{m,r}} \right\}_{m=1}^{{{N}_{T}}},\left\{ {{\mathbf{v}}_{m,r}} \right\}_{m=1}^{{{N}_{T}}}; \\ 
& \left\{ {{\mathbf{p}}_{j,r}} \right\}_{j=1}^{{{N}_{T}}L},\left\{ {{\mathbf{q}}_{m,r}} \right\}_{m=1}^{{{N}_{T}}},\left\{ {{\mathbf{d}}_{m,r}} \right\}_{m=1}^{{{N}_{T}}} \\ 
\end{aligned} \right) \\ 
& =2{{\mathbf{s}}_{T,r}}{{\mathbf{P}}_{r}}\left( {{{\mathbf{\tilde{S}}}}_{T}} \right)\mathbf{s}_{T,r}^{H}-2\mathbf{z}_{r}^{T}\left( {{{\mathbf{\tilde{s}}}}_{T}};{{{\mathbf{\tilde{S}}}}_{T}} \right){{\mathbf{s}}_{T,r}} \\ 
& +\sum\limits_{j=1}^{{{N}_{T}}L}{\left\{ \frac{{{\rho }_{1}}}{2}\left\| {{{\mathbf{\bar{E}}}}_{j,r}}\left( {{\mathbf{s}}_{T,r}}-{{\mathbf{s}}_{\operatorname{Ref},r}} \right)-{{\mathbf{h}}_{j,r}}+{{\mathbf{p}}_{j,r}} \right\|_{2}^{2} \right\}} \\ 
& +\sum\limits_{m=1}^{{{N}_{T}}}{\left\{ \frac{{{\rho }_{2}}}{2}\left\| {{\mathbf{\Sigma }}_{m,r}}{{\mathbf{s}}_{T,r}}-{{\mathbf{u}}_{m,r}}+{{\mathbf{q}}_{m,r}} \right\|_{2}^{2}+\frac{{{\rho }_{3}}}{2}\left\| {{{\mathbf{\bar{B}}}}_{m,r}}{{\mathbf{s}}_{T,r}}-{{\mathbf{v}}_{m,r}}+{{\mathbf{d}}_{m,r}} \right\|_{2}^{2} \right\}} \\ 
\end{aligned}
\end{equation}
where $\left\{ {{\mathbf{p}}_{j,r}} \right\}_{j=1}^{{{N}_{T}}L}$, $\left\{ {{\mathbf{q}}_{m,r}} \right\}_{m=1}^{{{N}_{T}}}$, and $\left\{ {{\mathbf{d}}_{m,r}} \right\}_{m=1}^{{{N}_{T}}}$ are Lagrange multipliers associated with the equality constraints ${{{\mathbf{\bar{E}}}}_{j,r}}\left( {{\mathbf{s}}_{T,r}}-{{\mathbf{s}}_{\operatorname{Ref,r}}} \right)={{\mathbf{h}}_{j}},j=1,2,...,{N_TL}$, $\left\| {{\mathbf{u}}_{m,r}} \right\|_{2}^{2}=\frac{1}{{{N}_{T}}}$, and ${{{\mathbf{\bar{B}}}}_{m,r}}{{\mathbf{s}}_{T,r}}={{\mathbf{v}}_{m,r}},m=1,2,...,{N_T}$, respectively. Here, ${{\rho }_{1}}$, ${{\rho }_{2}}$, and ${{\rho }_{3}}$ are penalty parameters.

\bibliographystyle{unsrt}
\bibliography{Refs}

\end{document}